\documentclass[english,aps, reprint, jcp, twocolumn]{revtex4}
\usepackage[L7x,T1]{fontenc}
\usepackage[utf8]{inputenc}
\usepackage{array}
\usepackage{bm}
\usepackage{amsmath}
\usepackage{amssymb}
\usepackage{graphicx}
\usepackage{esint}

\makeatletter

\newcommand{\noun}[1]{\textsc{#1}}
\providecommand{\tabularnewline}{\\}

\@ifundefined{textcolor}{}
{%
 \definecolor{BLACK}{gray}{0}
 \definecolor{WHITE}{gray}{1}
 \definecolor{RED}{rgb}{1,0,0}
 \definecolor{GREEN}{rgb}{0,1,0}
 \definecolor{BLUE}{rgb}{0,0,1}
 \definecolor{CYAN}{cmyk}{1,0,0,0}
 \definecolor{MAGENTA}{cmyk}{0,1,0,0}
 \definecolor{YELLOW}{cmyk}{0,0,1,0}
}

\makeatother

\usepackage{babel}
\begin{document}

\preprint{This line only printed with preprint option}

\title{Excitation transfer pathways in excitonic aggregates revealed by
the stochastic Schrödinger equation}

\author{Vytautas Abramavicius}

\email{vytautas.ab@gmail.com}

\affiliation{Vilnius University, Faculty of Physics, Department of Theoretical
Physics, Saulėtekio 9, LT-10222 Vilnius}

\author{Darius Abramavicius}

\email{darius.abramavicius@ff.vu.lt}

\affiliation{Vilnius University, Faculty of Physics, Department of Theoretical
Physics, Saulėtekio 9, LT-10222 Vilnius}
\begin{abstract}
We derive the stochastic Schrödinger equation for the system wave
vector and use it to describe the excitation energy transfer dynamics
in molecular aggregates. We suggest a quantum-measurement based method
of estimating the excitation transfer time. Adequacy of the proposed
approach is demonstrated by performing calculations on a model system.
The theory is then applied to study the excitation transfer dynamics
in a photosynthetic pigment-protein Fenna-Matthews-Olson (FMO) aggregate
using both the Debye spectral density and the spectral density obtained
from earlier molecular dynamics simulations containing strong vibrational
high-frequency modes. The obtained results show that the excitation
transfer times in the FMO system are affected by the presence of the
vibrational modes, however the transfer pathways remain the same.
\end{abstract}
\maketitle

\section{\label{sec:introduction}introduction}

Recent 2D spectroscopy studies of photosynthetic pigment-protein complexes
\cite{engel2007evidence,panitchayangkoon2010long} have shown the
evidence of coherent dynamics which may play a role in energy transfer
processes. These results sparked numerous debates whether the coherent
system dynamics are related to the observed high efficiency and speed
of the excitation energy transfer in such systems \cite{Novoderezhkin2010,Ishizaki2010,Olaya-Castro2011,C1CP20601J,Schlau-Cohen2011,Konig2012}.
Persistance of the coherent beats over picosecond and their robustness
contradict with predictions using conventional exciton relaxation
theory based on Markovian Redfield equation \cite{Pachon2011,Cheng2009}.
Possible vibronic contribution into some of these beats has been proposed
in a number of recent studies resulting in complex behavior of the
excitonic/vibronic 2D spectra \cite{mancal2012system,christensson2012origin,manvcal2010vibrational}.
Although long lasting beats in photosynthetic complexes reported by
Raman spectroscopy measurements are well known for a long time \cite{JRS:JRS1250260706},
the Raman experiments only provide information about the ground state
molecular. However, the coherent beats observed by 2D spectroscopy
have contributions from the electronic excited states and hence the
origin of the beats becomes obscure even in a such well-studied photosynthetic
complex as Fenna-Matthews-Olson (FMO) \cite{Milder2010,christensson2012origin,doi:10.1021/jp405421d}.
A number of experiments and theoretical studies has been recently
acomplished to disentangle the electronic/vibrational origin of these
beats in simple systems \cite{Butkus201240,Butkus201393,westenhoff2012coherent}. 

The strong interaction of molecular systems to environment greatly
increases the difficulty to theoretically describe the dynamics of
the systems because the environment has an infinite number of degrees
of freedom. Without the environment no such phenomena as relaxation
or energy transfer would be possible since only the macroscopic size
of the environment ensures truly irreversible dynamics of the system.
The most general way to calculate the quantum system dynamics is firstly
to solve the problem of the evolution of the whole closed quantum
system S+B, which can be described by its density operator $\widehat{\rho}\left(t\right)$,
and then to calculate the needed observables of the system S. If the
Hamiltonian $\widehat{H}$ characterizes this composite system, its
exact evolution in time is governed by the Liouville equation for
the full density operator $\widehat{\rho}\left(t\right)$:

\begin{equation}
i\frac{\textrm{d}}{\textrm{d}t}\widehat{\rho}\left(t\right)=\left[\widehat{\rho}\left(t\right),\widehat{H}\right].\label{eq:Liouv eq}
\end{equation}
Unfortunately practically it is not possible to solve it directly.
Therefore approximate methods which reduce the complexity of the open
quantum system are welcome \cite{valkunas2013molecular,breuer2002theory,may2004charge,Hashitsume1992}. 

Usually we are interested only in the dynamics of the system S which
we can describe using the reduced density operator $\widehat{\rho}_{red}\left(t\right)$.
It can be obtained by averaging over the environmental degrees of
freedom, i. e., performing the trace operation 
\begin{equation}
\widehat{\rho}_{\textrm{red}}\left(t\right)=\textrm{Tr}_{\textrm{B}}\left[\widehat{\rho}\left(t\right)\right].\label{eq:red dens op}
\end{equation}
Second order perturbation theory with respect to system-bath interaction
leads to the Redfield equation for the reduced density operator of
the system \cite{Redfield1957,valkunas2013molecular}. The Redfield
approach is sufficiently accurate and computationally effective for
rather large systems which interact weakly with the environment. The
closely related Lindblad equation \cite{Lindblad1976,Gorini1978}
method has the advantage of preserving the trace of the system reduced
density operator. The Lindblad equation can describe the system dynamics
at approximately the same level as Redfield equation \cite{:/content/aip/journal/jcp/133/6/10.1063/1.3458824}.
Time-adaptive Density Matrix Renormalization Group (t-DMRG) \cite{prior2010efficient,chin2011renormalisation},
Hierarchical Equations of Motion (HEOM) \cite{tanimura2006stochastic,ishizaki2009unified}
allow to incorporate the influence of the heat bath on the system
non-perturbatively. The HEOM method is formally exact for certain
types of environments, thus in principle it does not have restrictions
on the values of system parameters or system-bath coupling strength.
However in practice HEOM is computationally very costly, therefore
its application is limited only to small systems at sufficiently high
temperatures. The t-DMRG approach utilizes a unitary transformation
of the whole composite system into a linear 1D nearest-neighbor model.
It is formally exact and allows the usage of an arbitrary environmental
spectral density but it cannot include correlations between fluctuations
of different molecules, which may be important in spectroscopy \cite{:/content/aip/journal/jcp/134/17/10.1063/1.3579455}.
An iterative linearized density matrix (ILDM) stochastic approach
propagates the density matrix using the path integral technique for
the environment \cite{huo2010iterative}. Formally path integral is
exact approach as well, however a simplified version of ILDM allows
its practical application.

All the methods relying on the calculation of the reduced density
operator share a common property which allows only the investigation
of the statistically averaged behavior of the system S. However, in
this case the information about the instantaneous dynamic characteristics
of the system is lost. To investigate them the wavefunction description
of the system is preferrable. One of the approximate approaches for
the wavefunction is based on the smart guess of parametrized wavefunction
(so-called ansatz). A variational method is then used to determine
equations of motion for the wavefunction parameters \cite{chorosajev2013,Sun2010,Luo2010,Ye2012}
for the wavefunction to approximately satisfy the Schrödinger equation
with. However, the solution is restricted to the specific domain of
the ansatz. Alternative approaches for the wavefunction can be classified
as quantum jump methods and quantum state diffusion methods. The quantum
jump methods are based on a deterministic evolution of the system
wave vector with random jumps of the system state, e. g., surface
hopping when some vibrational adiabatic coordinate is explicitly included
\noun{\cite{dahlbom2002collective,beenken2002potential}} and they
govern the jump rates, or where jumps are realized by explicit jump
operators (so-called quantum Monte-Carlo approach) \cite{molmer1993monte,piilo2008non,piilo2009open,Rebentrost2009}.
Quantum state diffusion methods propagate the system wave vector under
the influence of continuous fluctuations which represent the action
of the environment \cite{gisin1992quantum,gisin1993quantum,diosi1998non,broadbent2012solving,zhong2013non}. 

In this paper we apply the stochastic approach to study the excitation
transfer times in molecular aggregates. Their histograms provide information
on excitation transport pathways. The main goal of this paper is to
investigate the dependency of the excitation energy transfer in the
FMO complex on the intra-molecular vibrations represented by the high
frequencies of the environmental spectral density. In Sec. \ref{sec:Theory}
we will see how this wave vector can be interpreted and then we derive
the general form of the stochastic Schrödinger equation (SSE). In
Sec. \ref{sec:Simulation-results} we define the procedure of transfer
time calculation and demonstrate its validity and the accuracy of
the SSE method on the simple dimer calculating the population dynamics
and transfer time distributions. Further we apply the SSE to study
the dynamics of the FMO complex and investigate the energy transfer
dependency on the intra-molecular vibrations in Sec. \ref{sec:FMO results}.

\section{\label{sec:Theory}Theory}

\subsection{\label{sub:Stochastic-Schr=0000F6dinger-equation}Stochastic Schrödinger
equation}

Let us consider a quantum system, defined by the Hamiltonian $\widehat{H}_{\textrm{S}}$.
In general the operator in some particular basis $\left|n\right\rangle $
can be represented in the following bra-ket form 
\begin{equation}
\widehat{H}_{\textrm{S}}=\overset{N}{\underset{n=1}{\sum}}\varepsilon_{n}\left|n\right\rangle \left\langle n\right|+\overset{N}{\underset{n\neq m}{\sum}}J_{nm}\left|n\right\rangle \left\langle m\right|,\label{eq:syst ham}
\end{equation}
where $N$ is the number of basis vectors (in the following we denote
them as sites), $\varepsilon_{n}$ - the energy of the $n$-th site,
$J_{nm}$ - the interaction energy between sites $n$ and $m$. The
environment is the harmonic heat bath of temperature $T$ which consists
of an infinite number of harmonic oscillators. Using the creation
- annihilation operators $\widehat{a}_{j}^{\dagger}$ and $\widehat{a}_{j}$
for the bath (the Planck's constant is set $\hbar=1$) we have
\begin{equation}
\widehat{H}_{\textrm{B}}=\underset{j}{\sum}\omega_{j}\widehat{a}_{j}^{\dagger}\widehat{a}_{j},\label{eq:bath ham}
\end{equation}
here $\omega_{j}$ is the frequency of the $j$-th oscillator. In
Eq. \eqref{eq:bath ham} the constant energy term is omitted because
it does not affect the dynamical properties of the system. The system
is linearly coupled to the environment via a set of system operators
$\widehat{L}_{n}$ and $\widehat{L}_{n}^{\dagger}$, thus the system
- bath interaction Hamiltonian will be written as 
\begin{equation}
\widehat{H}_{\textrm{SB}}=\kappa\underset{n}{\sum}\underset{j}{\sum}\left[\widehat{L}_{n}g_{nj}\widehat{a}_{j}^{\dagger}+h.c.\right],\label{eq:inter ham}
\end{equation}
where $h.c.$ denotes the Hermitian conjugate. Here the quantity $\kappa$
parametrizes the overall strength of the interaction between the system
and the environment, $g_{nj}$ are constants describing the coupling
strength between the $j$-th bath oscillator and the $n$-th system
operator $\widehat{L}_{n}$.

The composite system is closed, thus its state can be described by
the wave vector $\left|\Psi\left(t\right)\right\rangle $ which satisfies
the Schrödinger equation 
\begin{equation}
i\frac{\textrm{d}}{\textrm{d}t}\left|\Psi\left(t\right)\right\rangle =\left(\widehat{H}_{\textrm{S}}+\widehat{H}_{\textrm{B}}+\widehat{H}_{\textrm{SB}}\right)\left|\Psi\left(t\right)\right\rangle \equiv\widehat{H}\left|\Psi\left(t\right)\right\rangle .\label{eq:schrod eq}
\end{equation}
The solution of this equation formally can be written using the evolution
operator $\widehat{U}_{0}\left(t\right)$: 
\begin{equation}
\left|\Psi\left(t\right)\right\rangle =\widehat{U}_{0}\left(t\right)\left|\Psi\left(0\right)\right\rangle =\textrm{e}^{-i\widehat{H}t}\left|\Psi\left(0\right)\right\rangle .\label{eq:evol oper}
\end{equation}

Let us now switch to the interaction representation with respect to
the bath. In this representation we define a new time-dependent Hamiltonian
of the composite system 
\begin{equation}
\widehat{H}\left(t\right)\equiv\textrm{e}^{i\widehat{H}_{\textrm{B}}t}\left(\widehat{H}_{\textrm{S}}+\widehat{H}_{\textrm{SB}}\right)\textrm{e}^{-i\widehat{H}_{\textrm{B}}t}.\label{eq:time depend ham}
\end{equation}
Using the commutation relation of the bosonic creation - annihilation
operators $\left[\widehat{a}_{j},\widehat{a}_{k}^{\dagger}\right]=\delta_{jk}$
we can rewrite the Hamiltonian \eqref{eq:time depend ham} explicitly:
\begin{equation}
\widehat{H}\left(t\right)=\widehat{H}_{\textrm{S}}+\kappa\underset{n}{\sum}\underset{j}{\sum}\left[\widehat{L}_{n}g_{nj}\widehat{a}_{j}^{\dagger}\textrm{e}^{i\omega_{j}t}+h.c.\right].\label{eq:inter pict ham}
\end{equation}
The wave vector which is transformed to the interaction representation
$\left|\Psi\left(t\right)\right\rangle \rightarrow\left|\Psi\left(t\right)\right\rangle ^{(\textrm{I})}=\textrm{e}^{i\widehat{H}_{\textrm{B}}t}\left|\Psi\left(t\right)\right\rangle $
also satisfies the Schrödinger equation \eqref{eq:schrod eq} if we
substitute the Hamiltonian $\widehat{H}\rightarrow\widehat{H}\left(t\right)$.
Its solution is formally given by:
\begin{equation}
\left|\Psi\left(t\right)\right\rangle ^{(I)}=\widehat{U}\left(t\right)\left|\Psi\left(0\right)\right\rangle .\label{eq:evol op inter}
\end{equation}
Further on we will work only in the interaction representation, hence
for brevity we drop index $(I)$ above the wave vector. This wave
vector $\left|\Psi\left(t\right)\right\rangle $ of the composite
system encodes the full information about the evolution of both the
environment and the system. However, we are only interested in the
dynamics of the latter. 

Let us consider the initial state. We assume that initially the interaction
between the system and the environment is turned off and the system
is not correlated with the environment. The state of the system can
then be defined as $\left|\phi\right\rangle $. For the bath we must
have the thermal equilibrium, which is described by the canonical
density operator. Hence, the initial condition can be defined for
the density operator of the composite system as a tensor product \cite{zhong2013non,broadbent2012solving,diosi1998non,Strunz2005}:
\begin{equation}
\widehat{\rho}\left(0\right)=\left|\phi\right\rangle \left\langle \phi\right|\otimes\widehat{\rho}_{\textrm{B}}=\left|\phi\right\rangle \left\langle \phi\right|\otimes Z^{-1}\textrm{e}^{-\beta_{T}\widehat{H}_{\textrm{B}}},\label{eq:init dens op}
\end{equation}
where $\beta_{T}=1/T$ is the inverse temperature and $Z=\textrm{Tr}_{\textrm{B}}[\textrm{e}^{-\beta_{T}\widehat{H}_{\textrm{B}}}]$,
here the trace is over the bath degrees of freedom. 

Alternatively as all bath oscillators and the system are uncoupled
at $t=0$, the wave vector $\left|\Psi\left(0\right)\right\rangle $
of the composite system can be formally written in a chosen basis
for the bath $\left|\beta_{j}\right\rangle $ as a tensor product
as well 
\begin{equation}
\left|\Psi\left(0\right)\right\rangle =\left|\phi\right\rangle \otimes\left|\beta_{1}\right\rangle \otimes\left|\beta_{2}\right\rangle \otimes...\equiv\left|\phi\right\rangle \otimes\left|\bm{\beta}\right\rangle ,\label{eq:wave vect full init}
\end{equation}
where $\left|\beta_{i}\right\rangle $ describes the state of the
$i$-th oscillator. It is convenient to characterize the state of
the environmental oscillators using the coherent states of harmonic
oscillators (see Appendix \ref{sec:coher states}). The coherent states
basis are the Gaussian wavepackets which have the closest resemblence
with the classical description which is expected for the bath at high
temperatures. Additionally the coherent state $|\alpha\rangle$ has
very clear physical meaning: $\Re\alpha$ and $\Im\alpha$ are the
coordinate and the momentum expectation values of the oscillator,
respectively. As we show below we also simply obtain Gaussian stochastic
process using this basis set. 

By taking the scalar product of the full wave vector $\left|\Psi\left(0\right)\right\rangle $
and the vector $\left\langle \boldsymbol{\alpha}\right|=\left\langle \alpha_{1}\right|\otimes\left\langle \alpha_{2}\right|\otimes...$,
we find the wave vector of the system at the initial time 
\begin{equation}
\left|\psi\bm{\alpha}_{\phi\bm{\beta}}\left(0\right)\right\rangle =\textrm{e}^{-\bm{\alpha}^{\ast}\bm{\beta}}\left\langle \boldsymbol{\alpha}|\Psi\left(0\right)\right\rangle ,\label{eq:sys wave vect init}
\end{equation}
with respect to the bath states $|\alpha\rangle$ and $|\beta\rangle$.
Here $\bm{\alpha}^{*}\bm{\beta}=\underset{j}{\sum}\alpha_{j}^{*}\beta_{j}$.
At time $t$ we can then write:
\begin{eqnarray}
 &  & \left|\psi\bm{\alpha}_{\phi\bm{\beta}}\left(t\right)\right\rangle =\textrm{e}^{-\bm{\alpha}^{\ast}\bm{\beta}}\left\langle \boldsymbol{\alpha}|\Psi\left(t\right)\right\rangle \nonumber \\
 &  & \hspace{48bp}\equiv\textrm{e}^{-\bm{\alpha}^{\ast}\bm{\beta}}\left\langle \boldsymbol{\alpha}\left|\widehat{U}\left(t\right)\right|\boldsymbol{\beta}\right\rangle \left|\phi\right\rangle .\label{eq:sys wave vect t}
\end{eqnarray}
Recall that $\left|\phi\right\rangle $ does not involve the bath
state, hence, it has no indices $\alpha$ and $\beta$. 

At this point we can form the reduced density operator \eqref{eq:red dens op}
at an arbitrary time moment, which reads: 
\begin{eqnarray}
 &  & \widehat{\rho}_{\textrm{red}}\left(t\right)=\textrm{Tr}_{\textrm{B}}\left[\left|\Psi\left(t\right)\right\rangle \left\langle \Psi\left(t\right)\right|\right]\nonumber \\
 &  & =\int\frac{\textrm{d}^{2}\boldsymbol{\alpha}}{\pi}\textrm{e}^{-|\boldsymbol{\alpha}|^{2}}\left\langle \boldsymbol{\alpha}\right|\widehat{U}\left(t\right)\left|\phi\right\rangle \widehat{\rho}_{\textrm{B}}\left\langle \phi\right|\widehat{U}^{\dagger}\left(t\right)\left|\boldsymbol{\alpha}\right\rangle .\nonumber \\
\label{eq:red dens op coher}
\end{eqnarray}
Using the equilibrium bath density operator in the coherent state
basis (see Appendix \ref{sec:Equilibrium-density-operator}) in Eq.
\eqref{eq:red dens op coher} we obtain the system reduced density
operator expressed through the system wave vector: 
\begin{eqnarray}
 &  & \widehat{\rho}_{red}\left(t\right)=\int\frac{\textrm{d}^{2}\bm{\beta}}{\pi}\int\frac{\textrm{d}^{2}\boldsymbol{\alpha}}{\pi}\left[\prod_{j}p_{\alpha_{j}\beta_{j}}(\omega_{j})\right]\nonumber \\
 &  & \qquad\hspace{13bp}\times\left|\psi\bm{\alpha}_{\phi\bm{\beta}}\left(t\right)\right\rangle \left\langle \psi\bm{\alpha}_{\phi\bm{\beta}}\left(t\right)\right|,\label{eq:red dens op wave vect}
\end{eqnarray}
where 
\begin{eqnarray}
 &  & p_{\alpha\beta}(\omega_{j})=\bar{n}_{j}^{-1}\exp\left(-|\alpha|^{2}-|\beta|^{2}\textrm{e}^{\beta_{T}\omega_{j}}+\alpha^{*}\beta+\alpha\beta^{*}\right).\nonumber \\
\label{eq:prob distrib func}
\end{eqnarray}
We can see from Eq. \eqref{eq:red dens op wave vect} that $\widehat{\rho}_{\textrm{red}}\left(t\right)$
is given by the system wave vector defined in Eq. \eqref{eq:sys wave vect t}
and depends on the quantum variables of the environment $\alpha$
and $\beta$. These are complex-valued quantities representing a particular
configuration of the heat bath. Also notice that the temperature enters
only with respect to variable $\beta$. States $\left|\bm{\beta}\right\rangle $
are the \char`\"{}entry\char`\"{} states, which define the initial
thermal equilibrium density operator of the bath. This is the reason
why the thermal Boltzmann exponent is only related to variables $\beta$.
States $\left|\bm{\alpha}\right\rangle $ should be understood as
the \char`\"{}exit\char`\"{} states which are used to expand the final
state of the environment at an arbitrary time. 

However, the final expression can also be interpreted differently.
First, we notice that parameters $\alpha_{j}$ and $\beta_{j}$ are
continuous variables, which could be assumed as stochastic parameters.
Second, the factor $p_{\alpha\beta}(\omega_{j})$ has the form of
the probability density function of two variables $\alpha$ and $\beta$.
It follows that the operator $\left|\psi\bm{\alpha}_{\phi\bm{\beta}}\left(t\right)\right\rangle \left\langle \psi\bm{\alpha}_{\phi\bm{\beta}}\left(t\right)\right|$,
which is the matrix element of the density operator of the composite
system pure state, can be interpreted as the density operator of the
particular configuration of the system state $\psi$, with respect
to the bath stochastic configuration. Hence the system wave vector
$\left|\psi\bm{\alpha}_{\phi\bm{\beta}}\left(t\right)\right\rangle $,
can be interpreted as a stochastic system wave vector depending on
the particular configuration of the environment, characterized by
two stochastic complex-valued infinite-dimensional vectors $\bm{\alpha}$
and $\bm{\beta}$. Consequently, we can take one particular configuration
($\bm{\alpha}$,$\bm{\beta}$), calculate the vector $\left|\psi\bm{\alpha}_{\phi\bm{\beta}}\left(t\right)\right\rangle $
and the averaging of Eq. \eqref{eq:red dens op wave vect} with the
probability density in Eq. \eqref{eq:prob distrib func} necessarily
provides the proper reduced density matrix. It should be mentioned
that at this point the wave vector $\left|\psi\bm{\alpha}_{\phi\bm{\beta}}\left(t\right)\right\rangle $
is not normalized. 

The equation for the system wave vector $\left|\psi\bm{\alpha}_{\phi\bm{\beta}}\left(t\right)\right\rangle $
can be obtained by differentiating Eq. \eqref{eq:sys wave vect t}
with respect to time: 
\begin{eqnarray}
 &  & \frac{\textrm{d}}{\textrm{d}t}\left|\psi\bm{\alpha}_{\phi\bm{\beta}}\left(t\right)\right\rangle =\textrm{e}^{-\bm{\alpha}^{*}\bm{\beta}}\left\langle \boldsymbol{\alpha}\right|\frac{\textrm{d}}{\textrm{d}t}\widehat{U}\left(t\right)\left|\boldsymbol{\beta}\right\rangle \left|\phi\right\rangle \nonumber \\
 &  & =-i\widehat{H}_{\textrm{S}}\left|\psi\bm{\alpha}_{\phi\bm{\beta}}\left(t\right)\right\rangle -i\kappa\sum_{nj}\widehat{L}_{n}g_{nj}\textrm{e}^{i\omega_{j}t}\alpha_{j}^{*}\left|\psi\bm{\alpha}_{\phi\bm{\beta}}\left(t\right)\right\rangle \nonumber \\
 &  & -i\textrm{e}^{-\bm{\alpha}^{*}\bm{\beta}}\kappa\sum_{nj}\widehat{L}_{n}^{\dagger}g_{nj}^{*}\textrm{e}^{-i\omega_{j}t}\left\langle \boldsymbol{\alpha}\right|\widehat{a}_{j}\widehat{U}\left(t\right)\left|\boldsymbol{\beta}\right\rangle \left|\phi\right\rangle .\label{eq:sys vect deriv}
\end{eqnarray}
In the last term of Eq. \eqref{eq:sys vect deriv} we calculate the
quantity $\left\langle \boldsymbol{\alpha}\right|\widehat{a}_{j}\widehat{U}\left(t\right)\left|\boldsymbol{\beta}\right\rangle \left|\phi\right\rangle $
by writing the annihilation operator $\widehat{a}_{j}$ in the Heisenberg
representation $\widehat{a}_{j}\left(t\right)=\widehat{U}^{\dagger}\left(t\right)\widehat{a}_{j}\widehat{U}\left(t\right)$
leading to: $\widehat{a}_{j}\widehat{U}\left(t\right)=\widehat{U}\left(t\right)\widehat{a}_{j}\left(t\right)$.
Differentiating the operator $\widehat{a}_{j}\left(t\right)$ with
respect to time we obtain the equation: 
\begin{equation}
\frac{\textrm{d}}{\textrm{d}t}\widehat{a}_{j}\left(t\right)=-i\kappa\sum_{n}g_{nj}\textrm{e}^{i\omega_{j}t}\widehat{L}_{n}(t),\label{eq:Heis eq of motion}
\end{equation}
 with $\widehat{L}_{n}(t)=\widehat{U}^{\dagger}\left(t\right)\widehat{L}_{n}\widehat{U}\left(t\right)$
or 
\begin{equation}
\widehat{a}_{j}\left(t\right)=\widehat{a}_{j}-i\kappa\overset{t}{\underset{0}{\int}}\textrm{d}\tau\sum_{n}g_{nj}\textrm{e}^{i\omega_{j}\tau}\widehat{L}_{n}(\tau).\label{eq:annif op depend t}
\end{equation}
By using this result the third term of Eq. \eqref{eq:sys vect deriv}
becomes 
\begin{eqnarray}
 &  & \left\langle \boldsymbol{\alpha}\right|\widehat{U}\left(t\right)\widehat{a}_{j}\left(t\right)\left|\boldsymbol{\beta}\right\rangle \left|\phi\right\rangle =\beta_{j}\left|\psi\bm{\alpha}_{\phi\bm{\beta}}\left(t\right)\right\rangle \nonumber \\
 &  & -i\kappa\overset{t}{\underset{0}{\int}}\textrm{d}\tau\sum_{n}g_{nj}\textrm{e}^{i\omega_{j}\tau}\left\langle \boldsymbol{\alpha}\right|\widehat{U}\left(t\right)\widehat{U}^{\dagger}\left(\tau\right)\widehat{L}_{n}\widehat{U}\left(\tau\right)\left|\boldsymbol{\beta}\right\rangle \left|\phi\right\rangle \nonumber \\
\label{eq:third term}
\end{eqnarray}
and we can write Eq. \eqref{eq:sys vect deriv} in the following form:
\begin{eqnarray}
 &  & i\frac{\textrm{d}}{\textrm{d}t}\left|\psi\bm{\alpha}_{\phi\bm{\beta}}\left(t\right)\right\rangle =\widehat{H}_{\textrm{S}}\left|\psi\bm{\alpha}_{\phi\bm{\beta}}\left(t\right)\right\rangle \nonumber \\
 &  & +\kappa\sum_{n}\left[\widehat{L}_{n}z_{n}(t)+\widehat{L}_{n}^{\dagger}w_{n}(t)\right]\left|\psi\bm{\alpha}_{\phi\bm{\beta}}\left(t\right)\right\rangle \nonumber \\
 &  & -i\kappa^{2}\sum_{mn}\left[\widehat{L}_{n}^{\dagger}\overset{t}{\underset{0}{\int}}\textrm{d}\tau C_{nm}^{(0)}(t-\tau)\textrm{e}^{-\alpha^{*}\beta}\right.\nonumber \\
 &  & \times\left\langle \boldsymbol{\alpha}\right|\widehat{U}\left(t-\tau\right)\widehat{L}_{m}\widehat{U}\left(\tau\right)\left|\boldsymbol{\beta}\right\rangle \left|\phi\right\rangle \Biggr]\label{eq:stoch schro eq first}
\end{eqnarray}
In this equation we defined the following quantities: 
\begin{eqnarray}
 & {\displaystyle z_{n}(t)=\sum_{j}g_{nj}\alpha_{j}^{*}\textrm{e}^{i\omega_{j}t},}\label{eq:fluct zt def}\\
 & {\displaystyle w_{n}(t)=\sum_{j}g_{nj}^{*}\beta_{j}\textrm{e}^{-i\omega_{j}t},}\label{eq:fluct wt def}\\
 & C_{nm}^{(0)}(t)=\sum_{j}g_{nj}^{*}g_{mj}\textrm{e}^{-i\omega_{j}t}.\label{eq:zero T corr func}
\end{eqnarray}
Since according to previous discussion $\alpha_{j}$ and $\beta_{j}$
are stochastic complex quantities, $z_{n}\left(t\right)$ and $w\left(t\right)$
are Fourier transformations of these \textit{from the frequency domain
to the time domain}. This means that $z_{n}(t)$ and $w_{n}(t)$ are
complex-valued fluctuations. Let us calculate their correlation functions
$\mathcal{Z}_{nn}(t)=\left\langle z_{n}^{*}(t)z_{n}\left(0\right)\right\rangle _{\textrm{ens}}$
and $\mathcal{W}_{nn}(t)=\left\langle w_{n}^{*}(t)w_{n}\left(0\right)\right\rangle _{\textrm{ens}}$,
where $\left\langle ...\right\rangle _{\textrm{ens}}$ denotes the
statistical averaging operation using the Gaussian probability density
function from Eq. \eqref{eq:prob distrib func}. We find that 
\begin{eqnarray}
 &  & \mathcal{Z}_{nm}(t)=\sum_{j}\left(\overline{n}(\omega_{j})+1\right)g_{nj}^{*}g_{mj}\textrm{e}^{-i\omega_{j}t}\label{eq:eq:zt corr func}
\end{eqnarray}
and for $w_{n}(t)$: 
\begin{eqnarray}
 &  & \mathcal{W}_{nm}(t)=\sum_{j}\overline{n}(\omega_{j})g_{nj}g_{mj}^{\ast}\textrm{e}^{i\omega_{j}t}.\label{eq:wt corr func}
\end{eqnarray}
These functions depend on temperature, however as $T\to0$ we find
$\mathcal{Z}_{nm}(t)$ to be equivalent to Eq. \eqref{eq:zero T corr func},
hence, $C_{nm}^{(0)}(t)\equiv\mathcal{Z}_{nm}(t)|_{T=0}$. 

Eq. \eqref{eq:stoch schro eq first} can be denoted as the stochastic
Schrödinger equation. The first term on the right-hand side of the
equation with the Hamiltonian $\widehat{H}_{\textrm{S}}$ describes
the coherent evolution of the system. The second term accounts for
the influence of fluctuations $z_{n}(t)$ and $w_{n}(t)$ on the system.
The third term is related to the energy dissipation. The obtained
equation is not convenient due to the explicit dependency on the initial
system wave vector $\left|\phi\right\rangle $. Let us extract the
system wave vector from this term. Using Eq. \eqref{eq:non loc term}
and Eq. \eqref{eq:oper A defin} from Appendix \ref{sec:coher states}
we can obtain the most general form of the SSE for the system wave
vector:
\begin{eqnarray}
 &  & i\frac{\textrm{d}}{\textrm{d}t}\left|\psi\bm{\alpha}_{\phi\bm{\beta}}\left(t\right)\right\rangle =\widehat{H}_{\textrm{S}}\left|\psi\bm{\alpha}_{\phi\bm{\beta}}\left(t\right)\right\rangle \nonumber \\
 &  & +\kappa\sum_{n}\left[\widehat{L}_{n}z_{n}(t)+\widehat{L}_{n}^{\dagger}w_{n}(t)\right]\left|\psi\bm{\alpha}_{\phi\bm{\beta}}\left(t\right)\right\rangle \nonumber \\
 &  & -i\kappa^{2}\sum_{nm}\Biggl[\widehat{L}_{n}^{\dagger}\overset{t}{\underset{0}{\int}}\textrm{d}\tau C_{nm}^{(0)}(t-\tau)\widehat{A}_{\bm{\alpha}}\left(t-\tau\right)\widehat{L}_{m}\nonumber \\
 &  & \times\widehat{A}_{\boldsymbol{\alpha}}^{\dagger}\left(t-\tau\right)\Biggr]\left|\psi\bm{\alpha}_{\phi\bm{\beta}}\left(t\right)\right\rangle ,\label{eq:gen stoch schr eq}
\end{eqnarray}
Here $\widehat{A}_{\bm{\alpha}}\left(t-\tau\right)$ is the system
propagator when the state of the environment is $\bm{\alpha}$. Deriving
this result we did not make any approximations, hence it exactly describes
the evolution of the system with the Hamiltonian \eqref{eq:inter pict ham}.
We can see that Eq. \eqref{eq:gen stoch schr eq} has convolutionless
form, i. e. it is time-local, however, the evolution of the system
wave vector is non-Markovian because of the backward propagator $\widehat{A}_{\boldsymbol{\alpha}}^{\dagger}\left(t-\tau\right)$
acting on the wave vector. 

In the following we make the assumption that the action of the bath
is weak and we can then restrict ourselves only with the terms of
the order $\kappa^{2}$. In the expression \eqref{eq:gen stoch schr eq}
the non-local term is multiplied by the factor $\kappa^{2}$, thus
all functions inside the integral must be of the order $\kappa^{0}$.
This condition is satisfied when $\widehat{\mathcal{H}}_{\alpha}\left(t\right)\approx\widehat{H}_{\textrm{S}}$
leading to $\widehat{A}_{\bm{\alpha}}\left(t-\tau\right)\approx\exp\left(-i\widehat{H}_{\textrm{S}}\left(t-\tau\right)\right)$.
Additionally, as the initial parameters $\phi$, $\bm{\alpha}$ and
$\bm{\beta}$ now do not appear explicitly, we can drop them, i. e.,
$\left|\psi\bm{\alpha}_{\phi\bm{\beta}}\left(t\right)\right\rangle \equiv\left|\psi\left(t\right)\right\rangle $.
These simplifications turn the SSE into a simpler form: 
\begin{eqnarray}
 &  & i\frac{\textrm{d}}{\textrm{d}t}\left|\psi\left(t\right)\right\rangle =\widehat{H}_{\textrm{S}}\left|\psi\left(t\right)\right\rangle \nonumber \\
 &  & +\kappa\sum_{n}\left[\widehat{L}_{n}z_{n}(t)+\widehat{L}_{n}^{\dagger}w_{n}(t)\right]\left|\psi\left(t\right)\right\rangle \nonumber \\
 &  & -i\kappa^{2}\sum_{nm}\left[\widehat{L}_{n}^{\dagger}\overset{t}{\underset{0}{\int}}\textrm{d}\tau C_{nm}^{(0)}(\tau)\textrm{e}^{-i\widehat{H}_{\textrm{S}}\tau}\widehat{L}_{m}\textrm{e}^{i\widehat{H}_{\textrm{S}}\tau}\right]\left|\psi\left(t\right)\right\rangle .\nonumber \\
\label{eq:local stoch eq}
\end{eqnarray}
 The second term now introduces the fluctuations, while third term
takes care of the damping/dephasing .

We can note that the expressions of the SSE \eqref{eq:gen stoch schr eq}
and \eqref{eq:local stoch eq} resemble the form of the Redfield equation.
Consider the general form of the Redfield equation \cite{may2004charge,valkunas2013molecular}:
\begin{eqnarray}
 &  & \frac{\textrm{d}}{\textrm{d}t}\widehat{\rho}_{\textrm{red}}(t)=-i\left[\widehat{H}_{\textrm{int}},\widehat{\rho}_{\textrm{red}}\left(t\right)\right]-\left(\overset{t}{\underset{0}{\int}}\textrm{d}\tau\widehat{R}(\tau)\right)\widehat{\rho}_{\textrm{red}}(t),\nonumber \\
\label{eq:gen Redf eq}
\end{eqnarray}
where $\widehat{R}\sim\widehat{H}_{\textrm{int}}^{2}$ is a superoperator
responsible for the dissipation acting on the reduced system operator
$\widehat{\rho}_{\textrm{red}}$. Expression \eqref{eq:gen Redf eq}
is obtained using the same approximations are the SSE. Despite the
fact that the SSE has similar form and one could expect comparable
accuracy from both methods, the stochastic equation has one big advantage.
It is well-known that the Redfield equation leads to unphysical results
in certain regimes of parameters \cite{:/content/aip/journal/jcp/130/23/10.1063/1.3155214}.
The stochastic equation avoids this problem as the wavefunction can
be normalizeed at an arbitrary time and the final density matrix will
always be physical.

\subsection{\label{sub:actual model descr}Model with independent diagonal fluctuations}

In this work we investigate the stochastic dynamical characteristics
of molecular excitations in the aggregate consisting of $N$ molecules.
Each molecule is considered as a two-level system, characterized by
the excitation energy $\varepsilon_{n}$. We consider only a single
excitation in the aggregate, so state $|n\rangle$ denotes the excitation
residing on site $n$. It is often assumed that the interaction of
such system with the environment can be approximated by including
the diagonal fluctuations (to excitation energies) \cite{Rebentrost2009,rebentrost2009environment}.
In the stochastic equation we have to define operators $\widehat{L}_{n}$,
which couple the system with the environment, accordingly. For diagonal-only
fluctuations they become the projection operators $\widehat{L}_{n}=\left|n\right\rangle \left\langle n\right|$.
Thus, the fluctuations of the heat bath affect only the diagonal elements
of the system Hamiltonian $\widehat{H}_{\textrm{S}}$. Additionally
we assume that different projectors $\widehat{L}_{n}$ are coupled
to different sets of the bath oscillators \cite{Rebentrost2009,rebentrost2009environment}.
This makes the correlation functions $\mathcal{Z}_{nm}(t)$ and $\mathcal{W}_{nm}(t)$
diagonal. Taking that the environment of all sites is statistically
the same ($C_{nm}^{(0)}(\tau)\equiv\delta_{nm}C^{(0)}(\tau)$) Eq.
\eqref{eq:local stoch eq} for the system wave vector $\left|\psi\left(t\right)\right\rangle $
can then be written in the following way: 
\begin{eqnarray}
 &  & i\frac{\textrm{d}}{\textrm{d}t}\left|\psi\left(t\right)\right\rangle =\widehat{H}_{\textrm{S}}\left|\psi\left(t\right)\right\rangle +\kappa\sum_{n}\widehat{L}_{n}u_{n}(t)\left|\psi\left(t\right)\right\rangle \nonumber \\
 &  & -i\kappa^{2}\sum_{n}\left[\widehat{L}_{n}\overset{t}{\underset{0}{\int}}\textrm{d}\tau C^{(0)}(\tau)\textrm{e}^{-i\widehat{H}_{S}\tau}\widehat{L}_{n}\textrm{e}^{i\widehat{H}_{S}\tau}\right]\left|\psi\left(t\right)\right\rangle \nonumber \\
 &  & \equiv\left(\widehat{H}_{\textrm{S}}+\widehat{H}_{u}\left(t\right)\right)\left|\psi\left(t\right)\right\rangle ,\label{eq:schr eq with Ht}
\end{eqnarray}
where we have a new stochastic function $u_{n}\left(t\right)=z_{n}(t)+w_{n}(t)$.
The stochastic function $u_{n}\left(t\right)$ replaces functions
$z_{n}(t)$ and $w_{n}(t)$. Hence, the set of variables ($\bm{\alpha},\bm{\beta}$)
can now be replaced by a stochastic complex-valued functions of frequency
$u_{n}(\omega)$. The sole characteristics which fully defines $u_{n}(t)$
and $u_{n}(\omega)$ is the correlation function of $u$. Using Eq.
\eqref{eq:eq:zt corr func} and Eq. \eqref{eq:wt corr func} we find
\begin{eqnarray}
 &  & C_{nn}(t)=\sum_{j}\left[\left(\overline{n}(\omega_{j})+1\right)|g_{nj}|^{2}\textrm{e}^{-i\omega_{j}t}\right.\nonumber \\
 &  & \hspace{32bp}\left.+\overline{n}(\omega_{j})|g_{nj}|^{2}\textrm{e}^{i\omega_{j}t}\right].\label{eq:u_n corr func}
\end{eqnarray}

At this point it is convenient to introduce the spectral density of
the heat bath which describes the distribution of frequencies of the
environmental oscillators at $n$-th site. For our model we have 
\begin{equation}
C^{\prime\prime}\left(\omega\right)=\sum_{j}|g_{nj}|^{2}\delta\left(\omega-\omega_{j}\right).\label{eq:sp dens def}
\end{equation}
Extending it to negative frequencies we define $C^{\prime\prime}\left(-\omega\right)=-C^{\prime\prime}\left(\omega\right)$.
The correlation function is thus fully defined by the spectral density,
which is a continuous function of frequency for an infinite number
of bath oscillators \cite{valkunas2013molecular}: 
\begin{eqnarray}
 &  & C_{nn}(t)=\int\frac{\textrm{d}\omega}{2\pi}C^{\prime\prime}\left(\omega\right)\nonumber \\
 &  & \hspace{32bp}\times\left(\coth\frac{\omega\beta_{T}}{2}\cos\omega t-i\sin\omega t\right).\nonumber \\
\label{eq:full corr func}
\end{eqnarray}
Additionally, since $\mathcal{W}_{nm}(t)=0$ at zero temperature,
we have 
\begin{equation}
C_{nn}^{(0)}(\tau)=\int\frac{\textrm{d}\omega}{2\pi}C^{\prime\prime}\left(\omega\right)e^{-i\omega t},\label{eq:zero temp corr func}
\end{equation}
i. e., it is a Fourier transform of the spectral density.

A widely used model for the environment is based on the Debye spectral
density. Its form is an overdamped Lorentzian: 
\begin{equation}
C_{\textrm{Deb}}^{\prime\prime}\left(\omega\right)=\frac{2\lambda\omega\omega_{\textrm{D}}}{\omega^{2}+\omega_{\textrm{D}}^{2}},\label{eq:Debye sp dens}
\end{equation}
where $\lambda$ is the reorganization energy which characterizes
the system - bath coupling strength and $\omega_{\textrm{D}}\sim\tau_{\textrm{D}}^{-1}$
is the Debye frequency inversely proportional to the correlation time
of environmental fluctuations. The system - bath coupling strength
is defined via the parameter $\lambda$ (quantity $\kappa$ in Eq.
\ref{eq:schr eq with Ht} can be set to $1$). 

The SSE depends on the stochastic trajectory. Fluctuations having
a predefined correlation function can be generated using the Wiener
- Khinchin theorem in the frequency domain \cite{valkunas2013molecular}.
If the ergodicity condition is fulfilled the correlation function
can be defined by the Fourier transform of the stochastic trajectory:
\begin{eqnarray}
 &  & C(t)=\int\frac{\textrm{d}\omega}{2\pi}\textrm{e}^{-i\omega t}|u\left(\omega\right)|^{2}\label{eq:WK}
\end{eqnarray}
where 
\begin{equation}
u\left(\omega\right)=\int\textrm{d}t\textrm{e}^{i\omega t}u\left(t\right)
\end{equation}
is the Fourier transform of the stochastic trajectory $u\left(t\right)$.
Let us consider inverse procedure. To obtain $u\left(t\right)$ we
have to calculate $u\left(\omega\right)$ and then perform its inverse
Fourier transform. Since Eq. \eqref{eq:WK} is essentially a definition
of the Fourier transform, we notice that the function $u\left(\omega\right)$
is equal to:
\begin{equation}
A\left(\omega\right)=\textrm{e}^{i\varphi\left(\omega\right)}\sqrt{C\left(\omega\right)}.\label{eq:func A}
\end{equation}
We obtain $u\left(t\right)$ as a stochastic trajectory only when
we treat the phase $\varphi\left(\omega\right)$ as a stochastic function.
Itis essentially a random shift of the complex exponential in time.
Thus, the final expression of the noise $u\left(t\right)$ can be
written as
\begin{eqnarray}
 &  & u\left(t\right)=\int\frac{\textrm{d}\omega}{2\pi}\textrm{e}^{-i\omega t+i\varphi\left(\omega\right)}\sqrt{C^{\prime\prime}\left(\omega\right)\left(1+\coth\frac{\omega\beta_{T}}{2}\right)}.\nonumber \\
\label{eq:fluct express}
\end{eqnarray}
According to the central limit theorem the distribution of a sum obtained
from a large number of random variables is Gaussian. It follows that
the probability density function of the fluctuation \eqref{eq:fluct express}
remains Gaussian regardless of the distribution of the function under
the integral. For this reason we use the simple uncorrelated random
process to generate the function $\varphi\left(\omega\right)$ in
the interval $[0,2\pi)$. To have the real-value stochastic trajectory
of $u(t)$ at high temperature (classical fluctuations) we also set
$\varphi(\omega)=-\varphi(-\omega)$.

\section{\label{sec:Simulation-results}Simulation results}

\subsection{\label{sub:Population-relaxation-in tls}Population relaxation in
a two-level system }

One of the most widely used characteristics of the system dynamics
is the dependencies of the state populations on time. Averaged populations
in the site basis $\left|n\right\rangle $ can be calculated using
the wave vector, which is a $N$-component vector $\left|\psi\left(t\right)\right\rangle =\sum_{n}\psi_{n}(t)|n\rangle$.
The $n$-th population is then:
\begin{equation}
\rho_{nn}\left(t\right)=\left\langle \left|\psi_{n}\left(t\right)\right|^{2}\right\rangle _{\textrm{ens}},\label{eq:aver popul}
\end{equation}
where $\left\langle ...\right\rangle _{\textrm{ens}}$ denotes the
averaging over fluctuating trajectories. This quantity is essentially
a diagonal element of the density operator and thus can be readily
compared to other methods, e. g. Redfield or the Hierarchical Equations
of Motion (HEOM) approaches \cite{ishizaki2009unified,Cheng2005}. 

\begin{figure}[h]
\noindent \begin{centering}
\includegraphics[width=8cm]{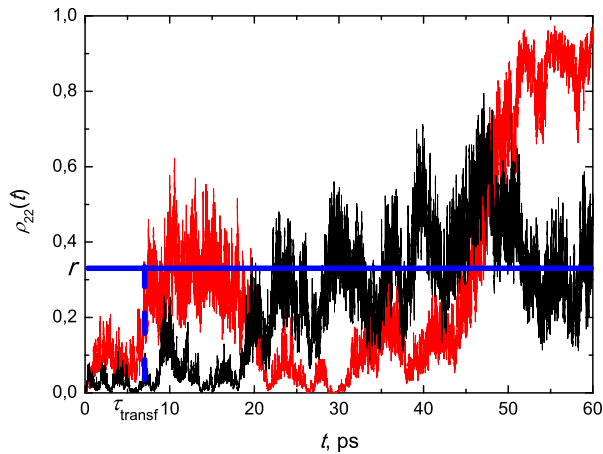}
\par\end{centering}

\caption{\label{fig:popul exp single traj} Single realizations of populations
$\rho_{22}(t)$ of two-level systems with coupling $J=4\textrm{ cm}^{-1}$
between the states calculated using Debye spectral density ($\omega_{D}=10\textrm{ fs}^{-1}$,
$\lambda=20\textrm{ cm}^{-1}$). Blue lines illustrate the procedure
of the excitation transfer time calculation.}
\end{figure}

To validate the theory let us consider relaxation properties in a
simple two-level system. Let's set the parameters of the model to
reflect weakly coupled two sites affected by small thermal noise.
So we set $\varepsilon_{1}=100\textrm{ cm}^{-1}$, $\varepsilon_{2}=0\textrm{ cm}^{-1}$,
the thermal noise is generated from the Debye spectral density with
short correlation time $\omega_{\textrm{D}}=10\textrm{ fs}^{-1}$,
with $\lambda=20\textrm{ cm}^{-1}$ and temperature $T=300\textrm{ K}$.
We next set the initial condition $|\psi\left(0\right)\rangle=\sum_{n}\delta_{n1}|n\rangle$.
By setting the intersite coupling to a small value ($J=4\textrm{ cm}^{-1}$)
in Fig. \ref{fig:popul exp single traj} we show two particular realizations
of the second site population $\left|\psi_{2}\left(t\right)\right|^{2}$
with respect to the fluctuating trajectory ($u_{1}(t)$ and $u_{2}(t)$).
Starting from the initial value $|\psi_{2}\left(0\right)|^{2}=0$
the population begins to rise in a stochastic fashion. Repeating the
same simulation for another realization of the noise we find initial
dynamics similar, but two curves quickly begin to diverge, thus reflecting
the decoherence process. 

\begin{figure}[h]
\noindent \begin{centering}
\includegraphics[width=8cm]{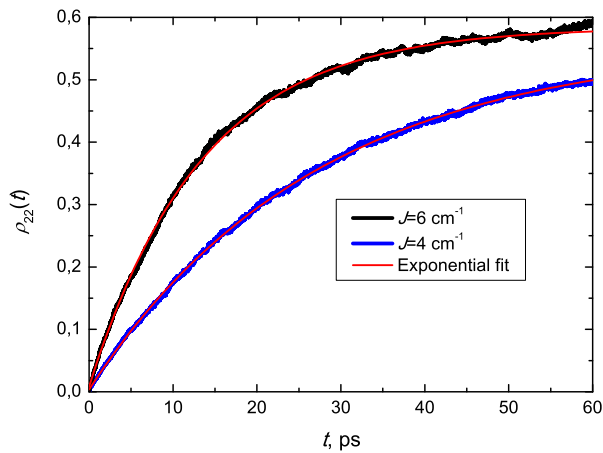}
\par\end{centering}

\caption{\label{fig:popul exp} Averaged populations $\rho_{22}(t)$ of two-level
systems with different coupling $J$ between the states calculated
using Debye spectral density ($\omega_{D}=10\textrm{ fs}^{-1}$, $\lambda=20\textrm{ cm}^{-1}$).
Parameters obtained from fitting when $J=4\textrm{ cm}^{-1}$: $A=0.56$,
$\tau=27.1\textrm{ ps}$; when $J=6\textrm{ cm}^{-1}$: $A=0.58$,
$\tau=13.4\textrm{ ps}$. Averaging is performed over $R=10000$ realizations.}
\end{figure}

Averaging such trajectories leads to the ensemble-averaged populations,
which is the ensemble-averaged density matrix. These are shown in
Fig. \ref{fig:popul exp} for two values of $J=4$ and $6$ $\textrm{cm}^{-1}$.
The averaged populations show exponential functional form, which is
confirmed by exponential fitting (parameters obtained from fitting:
for $J=4\textrm{ cm}^{-1}$, we get $A=0.56$, $\tau=27.1\textrm{ ps}$;
for $J=6\textrm{ cm}^{-1}$, we get $A=0.58$, $\tau=13.4\textrm{ ps}$)
. 

Indeed, in accord with the Fermi golden rule (FGR), which applies
in this weak coupling regime, the population $\rho_{22}(t)$ should
be approximated by the expression $\rho_{22}(t)=A\left[1-\exp(-t/\tau)\right]$
with $A=k_{1\rightarrow2}/(k_{1\rightarrow2}+k_{2\rightarrow1})$
and $\tau=1/(k_{1\rightarrow2}+k_{2\rightarrow1})$, where $k_{1\rightarrow2}$
and $k_{2\rightarrow1}$ are energy transfer rates from site 1 to
2 and vice versa, respectively. The values of these rates have been
obtained using the simple fitting, $k_{1\rightarrow2}=A/\tau$ and
$k_{2\rightarrow1}=(1-A)/\tau$. According to the FGR, the rates $k_{1\rightarrow2}$
and $k_{2\rightarrow1}$ must be proportional to $J^{2}$. This relation
is also confirmed investigating the results in Fig. \ref{fig:popul exp single traj}.
Hence in the weak intersite coupling limit, the SSE is consistent
with the FGR. 

\noindent 

\noindent 

It can be noticed that the populations do not exactly converge to
the values defined by the thermal distribution of system states. There
are two reasons for this. First, the exact Boltzmann equilibrium values
for the populations with respect to the level splittings of 100 cm$^{-1}$
can be obtained only when there is no interaction between the system
and the bath. Otherwise when system-bath coupling is on, the stationary
states become different and the equilibrium values in the simulations
are obtained with respect to the full system+bath Hamiltonian. Second,
the SSE is nevertheless approximate.

\subsection{\label{sub:Excitation-transfer-time}Excitation transfer time}

One of the main characteristic of the energy transfer is the transfer
time. This transfer time is the stochastic property, being unique
for each member of the ensemblel. Moreover this transfer time is a
stochastic property even for a single member of the ensemble. From
the theory of stochastic Markovian systems \cite{breuer2002theory}
it is known that the actual transfer time from the initial state to
the final one, when the process is characterized by a single rate
constant, must be a random number distributed according to the exponential
law with the properly defined mean transfer time, that is the inverse
of the rate, i. e. $p(t)=B\exp\left(-k_{1\rightarrow2}t\right)$.
Consequently the mean transfer time is given by $\tau_{1\to2}=1/k_{1\rightarrow2}$.
Hence, using the FGR (or the Redfield theory), the mean transfer time
can be evaluated as the inverse of calculated transfer rates. However,
the Redfield theory as well as the rate concepts are valid only for
the weak coupling regimes. In the case of intermediate or strong couplings,
the HEOM method allows to exactly propagate the density matrix, however,
the rates and the transfer times then become undefined. Additional
heuristic arguments may be necessary to define the transfer times
based on the density matrix population evolutions. We next devise
a stochastic method to simulate the excitation transfer time using
the SSE, which allows to properly define and evaluate the excitation
\emph{transfer time distribution function} even if it is not exponential. 

\begin{figure}[h]
\noindent \begin{centering}
\includegraphics[width=8cm]{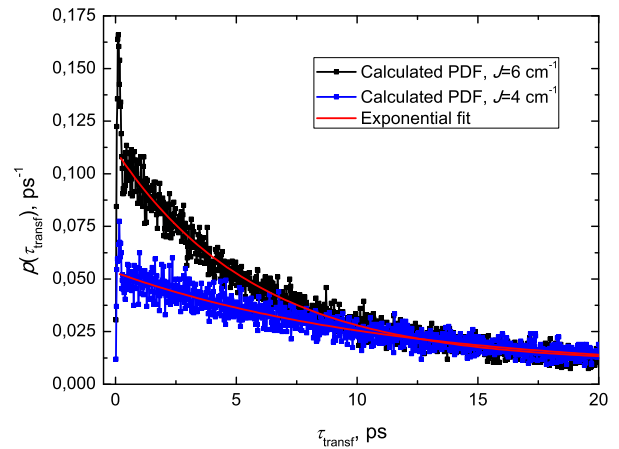}
\par\end{centering}

\caption{\label{fig:pdf exp}Probability density functions of the transfer
time $\tau_{\textrm{transf}}$ from the initially occupied state to
the unoccupied state in two-level systems with different coupling
$J$ between the states calculated using Debye spectral density ($\omega_{D}=10\textrm{ fs}^{-1}$,
$\lambda=20\textrm{ cm}^{-1}$). Parameters obtained from fitting
when $J=4\textrm{ cm}^{-1}$: $B=8.4\cdot10^{-5}$, $\tau_{1\to2}=11.1\textrm{ ps}$;
when $J=6\textrm{ cm}^{-1}$: $B=1.3\cdot10^{-4}$, $\tau_{1\to2}=5.7\textrm{ ps}$.}
\end{figure}

The meaning of the transfer time implies that we start with the predefined
state of the system and after some time we observe another state.
Hence, to calculate the transfer time to an arbitrary site of the
system, first we have to define the process of the measurement (detection)
of the excitation on the necessary site. This measurement procedure
can be constructed in the following way. The system wave vector with
components $\psi_{n}\left(t\right)$ is a stochastic variable depending
on a set of fluctuations $u_{k}\left(t\right)$ according to Eq. \eqref{eq:schr eq with Ht}.
Additionally, the magnitudes of the components of the wave vector
are generally nonzero. For this reason the system can be found in
an arbitrary state at an arbitrary time. We can \char`\"{}measure\char`\"{}
the excitation on an arbitrary site of the system by performing the
non-destructive quantum measurement of the system state. If the exciton
is detected on the $n$-th site, the system state collapses to $\left|n\right\rangle $,
we determine the arrival time and stop the propagation because the
state before arrival has now collapsed into a new state $\left|n\right\rangle $.
The statistical probabilities of these outcomes are $1-\left|\psi_{n}\left(t\right)\right|^{2}$
and $\left|\psi_{n}\left(t\right)\right|^{2}$, respectively. This
measurement process can be modeled using the Monte - Carlo method
by drawing a random number $r$ uniformly distributed in the interval
$[0,1)$ before starting the propagation (\ref{fig:popul exp single traj}).
Now during the propagation as soon as we find $r<\left|\psi_{n}\left(\tau\right)\right|^{2}$,
we \emph{register} the exciton on the $n$-th site and $\tau$ is
defined as the transfer time. Due to the fact that the system state
evolves stochastically and the random number $r$ takes unique values
for each propagation, the exciton detection condition is fulfilled
at different time moments in each realization. With a sufficient number
of realizations we can then calculate the distribution of the energy
transfer time. Hence, the algorithm of the exciton registration at
site $n$ and construction of the exciton transfer time distribution
can be summarized as follows:
\begin{enumerate}
\item A uniformly distributed random number $r$ is generated in the interval
$[0,1)$.
\item The system wave vector is propagated, and at every time step the condition
$r<\left|\psi_{n}\left(t\right)\right|^{2}$ is checked.
\item If the condition is satisfied, the propagation is stopped and the
exciton transfer time to the $n$-th site is recorded; otherwise the
propagation (step 2) is being continued.
\item 1-3 stages are repeated for the same system until a statistically
sufficient amount of results is obtained.
\item The distribution of transfer times is constructed as the histogram
of the arrival times.
\end{enumerate}
This procedure is illustrated in Fig. \ref{fig:popul exp single traj}:
the exciton detection time is marked by crossing point of $r$ and
$|\psi_{2}\left(t\right)|^{2}$. The corresponding distribution of
excitation transfer time from site $1$ to $2$ in the weakly-coupled
two-site system is presented in Fig. \ref{fig:pdf exp} As the FGR
holds in this case, we find proper exponential distribution of transfer
times. The mean values of the transfer time indeed correspond to the
transfer rates, determined from population evolution in Fig. \ref{fig:popul exp single traj}.

We must notice that the transfer time distributions in Fig. \ref{fig:pdf exp}
contain a sharp rise at short times which is not accounted by the
probabilistic theory of Markovian processes. This rise is the result
of transient processes caused by slight non-Markovianity of the bath
at short times originating from the finite-time correlation function
for the environmental fluctuations. In our case the correlation time
of the environment fluctuations is 10 fs. This initial rise corresponds
to this time. In the ideal Markovian case the fluctuation would be
infinitely fast (white noise) and the initial rise and transition
into the exponential function would happen at infinitesimal time interval. 

\noindent 

\subsection{\label{sub:Relaxation-in-str coupl dimer}Relaxation in a strongly-coupled
model system}

\begin{figure}[h]
\noindent \begin{centering}
\includegraphics[width=8cm]{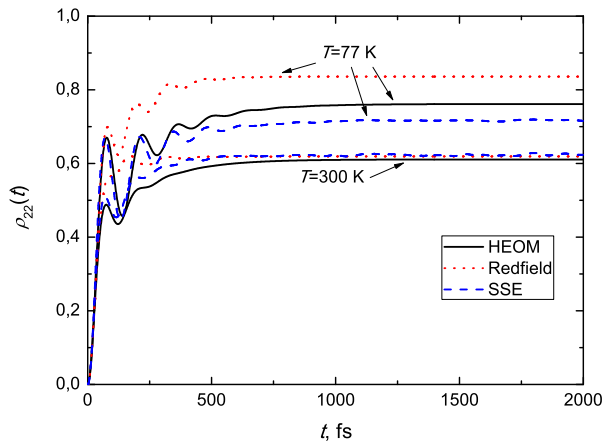}
\par\end{centering}

\caption{\label{fig:heom sse redf compar}Time dependencies of the population
$\rho_{22}\left(t\right)$ at different bath temperatures calculated
using SSE, Redfield and HEOM methods. For SSE averaging is performed
over $R=10000$ realizations.}
\end{figure}

We next consider the intermediate-to-strong coupling regime. Again
we study the two-site system, but we choose model parameters for intermediate
couplings consistent with ref. \cite{Ishizaki2010} : $\varepsilon_{1}=100\textrm{ cm}^{-1}$,
$\varepsilon_{2}=0\textrm{ cm}^{-1}$, $J_{12}=J_{21}=100\textrm{ cm}^{-1}$.
For the environment we choose the Debye spectral density (the same
for the both sites), with $\lambda=100\textrm{ cm}^{-1}$ and $\omega_{\textrm{D}}=100\textrm{ fs}^{-1}$
and study relaxation at two temperatures. The initial condition is
$|\psi\left(0\right)\rangle=\sum_{n}\delta_{n1}|n\rangle$. The population
of the second site is presented in Fig. \ref{fig:heom sse redf compar}.
We can see that at both heat bath temperatures ($T=300\textrm{ K}$
and $T=77\textrm{ K}$) the population rises very quickly in the first
100 fs and then performs oscillatory motion until it reaches the equilibrium
value. It should be noted that in case of higher environmental temperature
the amplitude of the population oscillations is smaller and they are
damped quicker ($\sim300\textrm{ fs}$) than in the case of low temperature
when oscillations die out after $\sim500\textrm{ fs}$. The oscillations
are mostly Rabi beats due to coupling $J$ and nonstationary initial
state. The damping is due to the bath. It has been discussed that
the approximate Redfield theory is not appropriate for this system,
since the relaxation rates and consequently the excitation transfer
times can not be accurately defined \cite{PhysRevB.84.245430}. For
comparison we also present the density matrix propagation results
using the Redfield theory with the time-dependent relaxation kernel
(Eq. \ref{eq:gen Redf eq}). The Redfield theory result shows large
deviations from the SSE result. However, as both are approximate,
the provided information is not sufficient to judge about correctness.
Therefore we additionally present the dependencies calculated using
the exact HEOM method. We can see that both methods give results that
coincide perfectly at the beginning of the simulation. We can also
notice that the equilibrium values of the populations calculated with
the SSE agree well with those obtained with the HEOM method and this
correspondence is better the higher is the temperature of the environment.
It is evident that the Redfield method gives largest errors. At low
temperatures some deviations between HEOM and SSE are moderate, however
the results of the stochastic method qualitatively reproduce the character
of the HEOM dependencies from short to intermediate times when transient
processes are present in the system.

\noindent 

\begin{figure}[h]
\noindent \begin{centering}
\includegraphics[width=8cm]{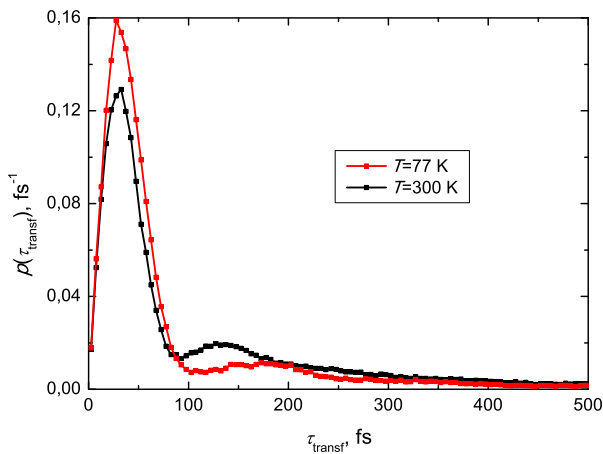}
\par\end{centering}

\caption{\label{fig:strong coup transf time distr}Probability density functions
of the transfer time $\tau_{\textrm{transf}}$ from the initially
occupied state to the unoccupied state in the two-level system at
different bath temperatures calculated using Debye spectral density
($\omega_{\textrm{D}}=100\textrm{ fs}^{-1}$, $\lambda=100\textrm{ cm}^{-1}$). }
\end{figure}

The dynamics of the two-site system with the above-given parameters
and heat bath characteristics should be rather non-Markovian due to
the long decay time of the bath correlation function (100 fs) and
this behavior must be reflected in the distributions of the exciton
transfer time. These distributions are presented in Fig. \ref{fig:strong coup transf time distr}.
Comparing these results with distributions obtained for the weakly
interacting Markovian system (Fig. \ref{fig:pdf exp}) we clearly
see that now the distributions are not exponential which indicates
the significance of non-Markovian effects in this two-site system.
The second peak in the transfer time signifies the coherent components.
We can notice that the duration of the initial rise of the probability
density functions corresponds to the relaxation time of the heat bath
($\omega_{\textrm{D}}=100\textrm{ fs}^{-1}$). Such non-exponential
distributions do not correspond to any process described by simple
constant-rate equations, which define the transfer mean times. Thus,
in this respect the SSE approach has an advantage in describing energy
transfer over the density matrix approaches.

\noindent 

\section{\label{sec:FMO results}Case study: Exciton dynamics in the FMO complex}

\begin{figure}[h]
\noindent \begin{centering}
\includegraphics[width=8cm]{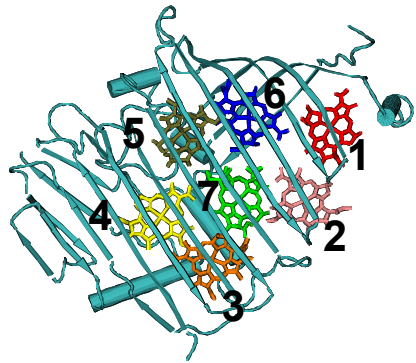}
\par\end{centering}

\caption{\label{fig:fmo structure}Arrangement of the bacteriochlorophylls
in one monomer of the FMO complex.}
\end{figure}

FMO complex found in the green sulfur bacteria is the first pigment
- protein which had its structure revealed using the method of X-ray
crystallography, hence it is one of the best studied photosynthetic
aggregates \cite{engel2007evidence,mohseni2008environment,panitchayangkoon2010long,Milder2010}.
FMO complex is a trimer consisting of 3 identical monomers which are
formed from 8 bacteriochlorophyll (BChl) molecules supported by a
rigid protein carcass (Fig. \ref{fig:fmo structure}). In green sulfur
bacteria the FMO aggregate acts as a molecular wire which transports
the excitation energy from the light-harvesting chlorosomes to the
reaction centers of the I type located in the membrane \cite{engel2007evidence,mohseni2008environment,panitchayangkoon2010long}.
We next apply the the SSE theory and the simulation protocol described
above to study the energy transfer dynamics in the FMO aggregate.

\noindent 

The FMO complex is described by the Hamiltonian adapted from previous
publications \cite{mohseni2008environment}. We assume that the FMO
system consists only of 7 sites corresponding to different BChl molecules.
The 8th molecule is not taken into account due to its weak coupling
with the rest of the BChls. Setting the energy of the 3rd site, through
which the energy excitation travels to the reaction center, to zero,
we obtain the matrix:

\begin{table}[h]
\begin{centering}
\begin{tabular}{c|>{\centering}m{1cm}|>{\centering}m{1cm}|>{\centering}m{1cm}|>{\centering}m{1cm}|>{\centering}m{1cm}|>{\centering}m{1cm}|>{\centering}m{1cm}|}
\multicolumn{1}{c}{} & \multicolumn{1}{>{\centering}m{1cm}}{1.} & \multicolumn{1}{>{\centering}m{1cm}}{2.} & \multicolumn{1}{>{\centering}m{1cm}}{3.} & \multicolumn{1}{>{\centering}m{1cm}}{4.} & \multicolumn{1}{>{\centering}m{1cm}}{5.} & \multicolumn{1}{>{\centering}m{1cm}}{6.} & \multicolumn{1}{>{\centering}m{1cm}}{7.}\tabularnewline
\cline{2-8} 
1. & \vspace{5bp}
280\vspace{5bp}
 & \vspace{5bp}
-106\vspace{5bp}
 & \vspace{5bp}
0\vspace{5bp}
 & \vspace{5bp}
0\vspace{5bp}
 & \vspace{5bp}
0\vspace{5bp}
 & \vspace{5bp}
-8\vspace{5bp}
 & \vspace{5bp}
-4\vspace{5bp}
\tabularnewline
\cline{2-8} 
2. &  & \vspace{5bp}
420\vspace{5bp}
 & \vspace{5bp}
28\vspace{5bp}
 & \vspace{5bp}
0\vspace{5bp}
 & \vspace{5bp}
0\vspace{5bp}
 & \vspace{5bp}
13\vspace{5bp}
 & \vspace{5bp}
0\vspace{5bp}
\tabularnewline
\cline{2-8} 
3. &  &  & \vspace{5bp}
0\vspace{5bp}
 & \vspace{5bp}
-62\vspace{5bp}
 & \vspace{5bp}
0\vspace{5bp}
 & \vspace{5bp}
0\vspace{5bp}
 & \vspace{5bp}
17\vspace{5bp}
\tabularnewline
\cline{2-8} 
4. &  &  &  & \vspace{5bp}
175\vspace{5bp}
 & \vspace{5bp}
-70\vspace{5bp}
 & \vspace{5bp}
-19\vspace{5bp}
 & \vspace{5bp}
-57\vspace{5bp}
\tabularnewline
\cline{2-8} 
5. &  &  &  &  & \vspace{5bp}
320\vspace{5bp}
 & \vspace{5bp}
40\vspace{5bp}
 & \vspace{5bp}
-2\vspace{5bp}
\tabularnewline
\cline{2-8} 
6. &  &  &  &  &  & \vspace{5bp}
360\vspace{5bp}
 & \vspace{5bp}
32\vspace{5bp}
\tabularnewline
\cline{2-8} 
7. &  &  &  &  &  &  & \vspace{5bp}
260\vspace{5bp}
\tabularnewline
\cline{2-8} 
\end{tabular}
\par\end{centering}

\caption{\label{tab:FMO-Hamiltonian.}Matrix elements of the FMO Hamiltonian
given in $\textrm{cm}^{-1}$. Sites are numbered according to the
crystallographic nomenclature \cite{fenna1977atomic}.}
\end{table}

In all simulations of the FMO system the initial state was chosen
to be a superposition $\left|\psi\left(0\right)\right\rangle =\frac{1}{\sqrt{2}}\left(\left|1\right\rangle +\left|6\right\rangle \right)$.
This state is chosen because the 1st and the 6th BChl molecules are
nearest to the light-harvesting chlorosomes where the excitation is
created \cite{mohseni2008environment}. The interaction with the environment
induces fluctuations of the excitation energies of the BChl molecules.
Classical correlation functions of these energy fluctuations for every
BChl molecule have been estimated by the Olbrich et al molecular dynamics
(MD) simulations of the whole FMO complex in the solution \cite{olbrich2011theory}.
These correlation functions have been approximated by a combination
of exponents and decaying oscillations. After performing the Fourier
transformation the spectral densities of different BChl molecules
at room temperature ($T=300$ K) have been obtained: 
\begin{eqnarray}
 &  & C_{\textrm{MD},n}^{\prime\prime}\left(\omega\right)=\frac{2}{\pi\hbar}\tanh\left(\frac{\beta_{T}\hbar\omega}{2}\right)\left[\overset{N_{0}}{\underset{m=1}{\sum}}\frac{\eta_{n,m}\gamma_{n,m}}{\gamma_{n,m}^{2}+\omega^{2}}\right.\nonumber \\
 &  & \hspace{10bp}\hspace{8bp}\hspace{30bp}\left.+\overset{N_{0}}{\underset{m=1}{\sum}}\frac{\widetilde{\eta}_{n,m}\widetilde{\gamma}_{n,m}}{2\left(\widetilde{\gamma}_{n,m}^{2}+\left(\omega-\widetilde{\omega}_{n,m}\right)^{2}\right)}\right].\nonumber \\
\label{eq:MD sp dens}
\end{eqnarray}
$\eta_{n}$, $\gamma_{n}$, $\widetilde{\eta}_{n}$ and $\widetilde{\omega}_{n}$
are parameters best fitting the corresponding correlation functions
and $N_{0}$ is the number of terms in the sum. In this expression
the factor $\tanh\left(\frac{\beta_{T}\hbar\omega}{2}\right)$ is
introduced to take into account the temperature dependence of the
parameters. In the following we denote this spectral density as the
MD spectral density.

\begin{figure}[h]
\noindent \begin{centering}
\includegraphics[width=8cm]{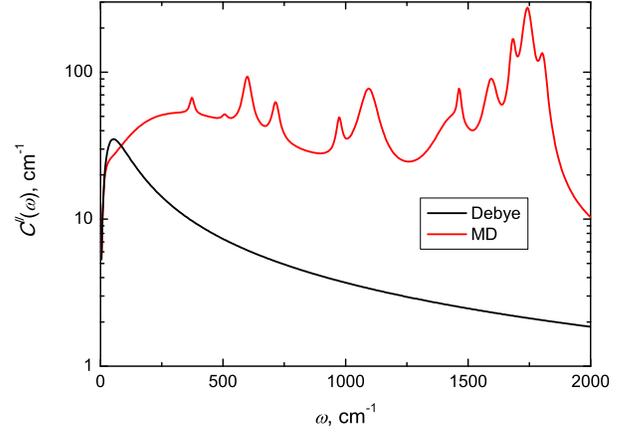}
\par\end{centering}

\caption{\label{fig:debye olbr sp dens}Debye and MD spectral densities used
for FMO complex. Parameters of Debye spectral density: $\omega_{D}=100\textrm{ fs}^{-1}$,
$\lambda=35\textrm{ cm}^{-1}$ were set according to Ref. \cite{ishizaki2009theoretical}.}
\end{figure}

The spectral density given above consists mainly of two terms. The
first is a Debye term determining the overdamped low frequency modes.
The second part reflects the high-frequency modes. These should be
associated with the intra-molecular vibrations. As the intra-molecular
vibrational frequencies are the same for all chrolophylls, while their
amplitudes vary from site to site, for simplicity we assume the averaged
spectral density for all BChl molecules with $N_{0}=12$ terms in
Eq. \eqref{eq:MD sp dens}. As a reference, a model without high frequency
intra-molecular vibrations based on the Debye spectral density (Eq.
\eqref{eq:Debye sp dens}) is used as well. Both Debye and MD spectral
densities have similar low-frequency part, while they are different
at high frequencies as shown in Fig. \ref{fig:debye olbr sp dens}.
The low frequency part also corresponds to the experimentally determined
spectral density \cite{wendling2000electron} in this range of frequencies.

\noindent 

\begin{figure}[h]
\noindent \begin{centering}
\includegraphics[width=8cm]{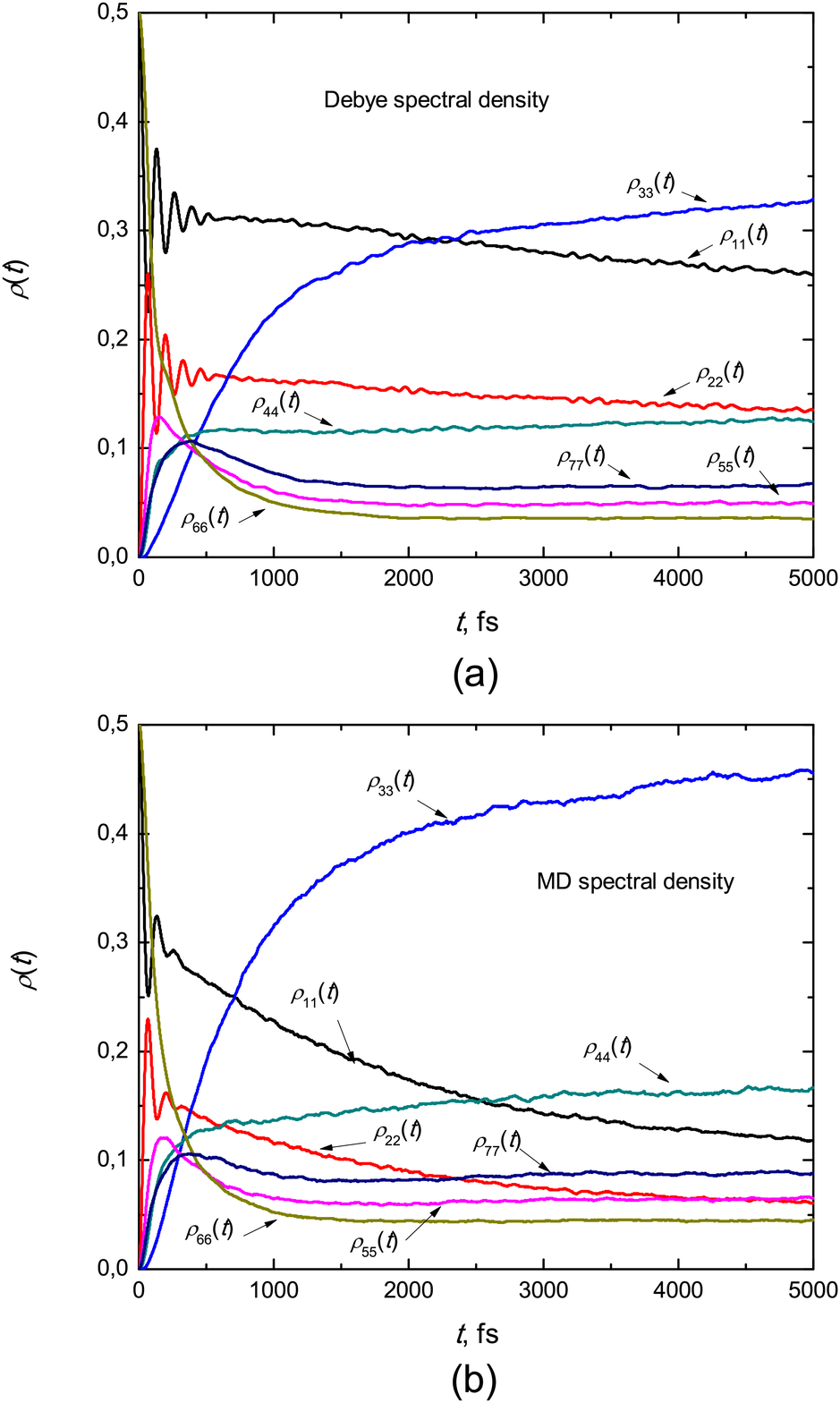}
\par\end{centering}

\caption{\label{fig:popul site basis}System population dependencies on time
calculated in the site basis: (a) with Debye spectral density ($\omega_{\textrm{D}}=100\textrm{ fs}^{-1}$,
$\lambda=35\textrm{ cm}^{-1}$), (b) with MD spectral density. Averaging
is performed over $R=10000$ realizations.}
\end{figure}

The dynamics and relaxation of the excitation in the aggregate can
be investigated by first analyzing time dependencies of the site populations.
These, averaged over the ensemble according to Eq. \ref{eq:aver popul}
at the temperature $T=300$, are presented in Fig. \ref{fig:popul site basis}.
We can see from both figures that when the system approaches equilibrium
the value of the population $\rho_{33}\left(t\right)$, which corresponds
to the site with the lowest energy $\varepsilon_{3}=0$, becomes the
largest in accord with previous simulations \cite{Nalbach2012}. Equilibrium
values of populations of other sites are also ordered in accord with
their energy corresponding to proper thermal equilibrium. Time dependencies
of the populations calculated with Debye spectral density (Fig. \ref{fig:popul site basis}
(a)) show that despite of populations $\rho_{44}\left(t\right)$,
$\rho_{55}\left(t\right)$, $\rho_{66}\left(t\right)$ and $\rho_{77}\left(t\right)$
reaching equilibrium after $\sim2\textrm{ ps}$, other curves are
still not stationary, thus the full relaxation of the system occurs
in more than 5 ps. The results obtained with MD spectral density are
depicted in Fig. \ref{fig:popul site basis} (b) demonstrate qualitatively
similar but slightly quicker relaxation process. The coherent evolution
as well as delocalized excitons in the system can be recognized from
the oscillations of the presented populations. From the calculations
with Debye spectral density we can see that oscillations decay after
$\sim500\textrm{ fs}$. Using MD spectral density we obtain smaller
amplitudes of oscillations and they decay faster - after $\sim300\textrm{ fs}$.
That manifests the decay of electronic coherences. Hence the MD spectral
density seems to slightly speed-up the relaxation dynamics without
noticeable qualitative differences.

\begin{figure}[h]
\noindent \begin{centering}
\includegraphics[width=8cm]{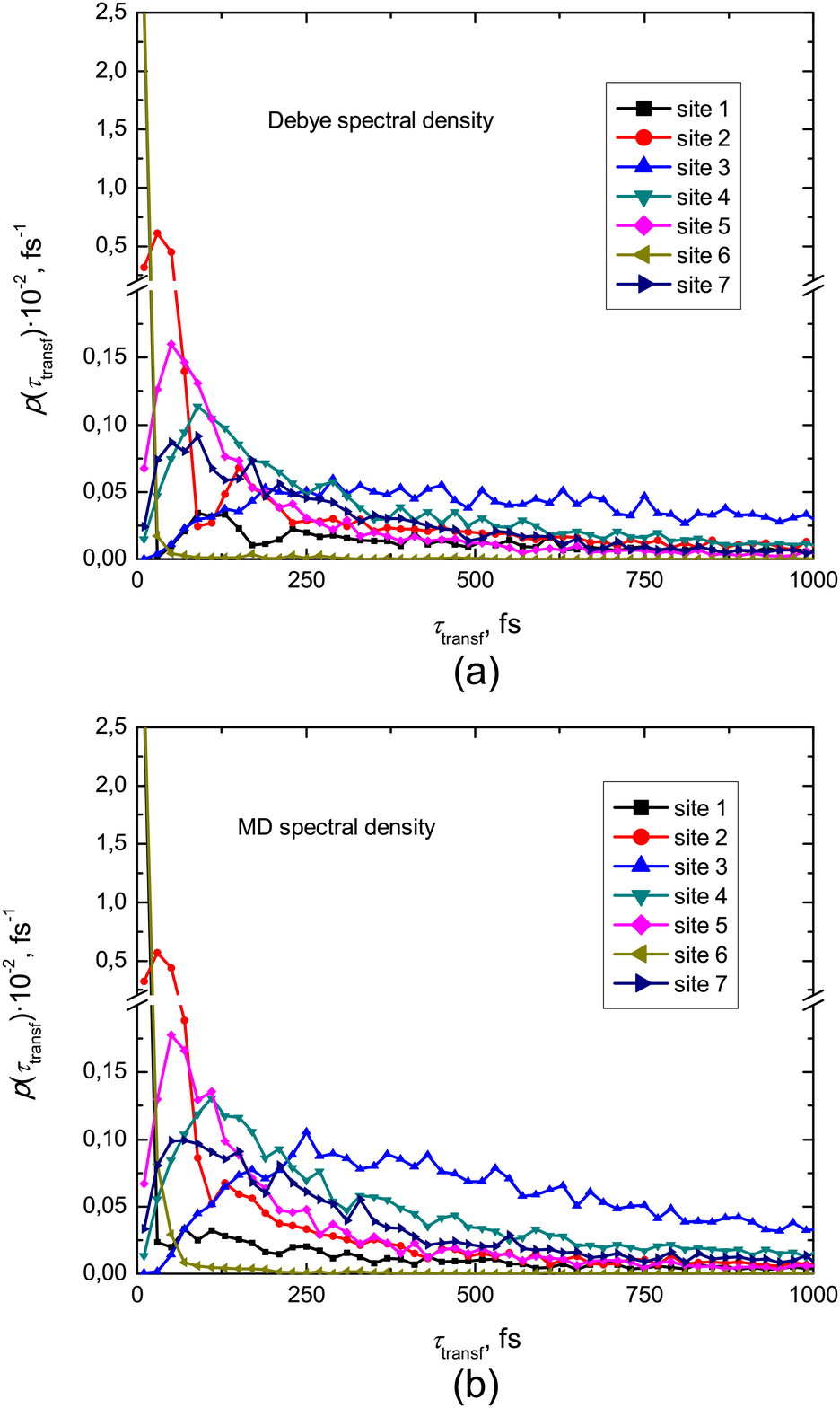}
\par\end{centering}

\caption{\label{fig:transf time all}Distributions of energy excitation transfer
time calculated for all sites of the system: (a) with Debye spectral
density ($\omega_{\textrm{D}}=100\textrm{ fs}^{-1}$, $\lambda=35\textrm{ cm}^{-1}$),
(b) with MD spectral density.}
\end{figure}

The most important result of this study which is available from the
SSE is the distribution of excitation transfer times. These distributions
for all seven sites calculated at temperature $T=300$ K are presented
in Fig. \ref{fig:transf time all}. It is evident that all distributions
calculated with both spectral densities greatly differ from the exponential
form. We can notice that with both spectral densities result in similar
overall arrangement of the exciton transfer time distribution curves
and also the most probable transfer times. It is evident that the
energy excitation can be registered at the 1st or the 6th site in
the shortest time. Analyzing the positions of the maxima of the probability
distributions we can see that the exciton travels through the FMO
complex in such order: 2nd site, 5th site, 7th site, 4th site and
it takes the longest time for the exciton to arrive at the 3rd site.
Hence, the transfer time distributions reveal the excitation transfer
pathways in multi-site excitonic systems. From the Fig. \ref{fig:transf time all}
we can also see that the transfer time distribution at the 3rd site
is the broadest, which means that in this case the time $\tau_{\textrm{transf}}$
has the biggest uncertainty. 

\section{\label{sec:Discussion}Discussion}

In this paper we employ the SSE to study relaxation of molecular excitations
where the system state is described by the stochastic vector $\left|\psi\left(t\right)\right\rangle $.
While the mathematical formulation of the problem is transparent,
the physical meaning of stochastic quantities requires additional
discussion. The origin of the stochasticity of $|\psi(t)\rangle$
vector stems from the fact that the environmental quantum variables,
$\bm{\alpha}$ and $\bm{\beta}$, are interpreted as complex-valued
Gaussian fluctuations acting on our investigated system. This step
introduces the ambiguity in interpretation of the stochastic vector
$|\psi(t)\rangle\equiv\left|\psi\bm{\alpha}_{\phi\bm{\beta}}\left(t\right)\right\rangle $. 

One's initial guess might be that the wavevector describes the state
of the reduced system $\psi$ at an arbitrary time. This initial guess
contains a flaw since based on the quantum mechanics the system and
bath parts inevitably must be entangled as time passes and the \char`\"{}reduced
wavefunction of the system\char`\"{} does not exist. The state of
the composite system+bath system must be understood as a linear superposition
in the basis of composite system+bath vectors $|n\bm{\alpha}\rangle$:
\begin{equation}
|\Psi\rangle=\sum_{n}\int\frac{\mathrm{d}\bm{\alpha}}{\pi}c_{n\bm{\alpha}}(t)|n\bm{\alpha}\rangle;
\end{equation}
 $n$ is taken as a discrete variable and $\bm{\alpha}$ is a continuous,
while $c_{n\bm{\alpha}}(t)$ is the specific component with respect
to the basis. What does $\left|\psi\bm{\alpha}_{\phi\bm{\beta}}\left(t\right)\right\rangle $
stand for? Let us remind that this quantity was obtained by taking
the factorized initial condition $\left|\phi\bm{\beta}\right\rangle $,
which properly denotes the initial state. The state is then propagated
for a time $t$ by an evolution operator. $\left|\psi\bm{\alpha}_{\phi\bm{\beta}}\left(t\right)\right\rangle $
is thus a \emph{specific} $\bm{\alpha}$\emph{ component} \emph{of
the complete wavefunction} with respect to the initial condition $\left|\phi\bm{\beta}\right\rangle $.
This component becomes available by collapsing the environment wavefunction
into the $|\bm{\alpha}\rangle$ state or in other words by measurement
of the $|\bm{\alpha}\rangle$ state. Probability of such collapse,
which should be realized by a quantum measurement, is given by the
norm of the vector $\left|\psi\bm{\alpha}_{\phi\bm{\beta}}\left(t\right)\right\rangle $.
Hence, the SSE describes a single $\bm{\alpha}$ component of the
full wavefunction. So how about other components and where does the
stochasticity comes from? This is an additional ingredient in the
stochastic equation, which takes into account the probability density
(Eq. \eqref{eq:prob distrib func}). Notice that the complete specification
of vectors $\bm{\alpha}$ and $\bm{\beta}$ completely determines
functions $z_{n}(t)$ and $w_{n}(t)$. These functions are merely
kind-of Heisenberg representation of $\bm{\alpha}$ and $\bm{\beta}$.
So when the component of the complete wavefunction is chosen, the
whole \char`\"{}trajectories\char`\"{} $z_{n}(t)$ and $w_{n}(t)$
are specified and stochasticity is removed. The different component
is obtained by taking $\bm{\alpha}'\neq\bm{\alpha}$, hence, by taking
another function $z_{n}'(t)$ corresponding to $\bm{\alpha}'$. Smart
way of choosing set of different $\bm{\alpha}$s (or $u_{n}(t)$ as
described in Sec. \ref{sub:actual model descr}) guarantees proper
probability distribution of different trajectories. Consequently the
whole wavefunction of the composite system is calculated by combining
the ensemble of functions $z_{n}(t)$ reflecting all possible realizations
of vector $\bm{\alpha}$. In this sense the whole procedure of averaging
over various trajectories becomes similar to the path-integral formulation.
The paths are distributed non-uniformly, but follow the Gaussian distribution.

Now the natural question is whether some interpretation of a single
$z(t)$ and $w(t)$ trajectory is possible? We would like to argue
that\emph{ yes} \emph{it is}. Recall that a vector characterizing
the system of interest alone can be defined if we measure the state
of the environment. Thus, after performing this measurement at $\bm{\alpha}$
the full wavefunction of the composite system becomes a tensor product
- $\left|\Psi\right\rangle =|\psi\bm{\alpha}\rangle$. The probabilities
of the site populations should be understood also conditioned on the
state $\bm{\alpha}$. These conditional probabilities thus reflect
the renormalization procedure, which we implied before. The $|\psi\bm{\alpha}\rangle$
vector is then the system wave vector conditioned on the result of
the measurement of the environment state at $\bm{\alpha}$ (also,
provided the initial state was $\bm{\beta}$). According to the SSE
(Eq. \eqref{eq:gen stoch schr eq}) the $\kappa^{1}$ term induces
the fluctuations of the Hamiltonian elements. Hence, a single configuration
of the bath $(\bm{\alpha},\bm{\beta})$ defines the full trajectory
of the fluctuating Hamiltonian, or the effect of the bath configuration
$(\bm{\alpha},\bm{\beta})$ on the whole time interval through functions
$z(t)$ and $w(t)$. As the bath configuration has the probability
$\exp\left(-|\alpha|^{2}-|\beta|^{2}\textrm{e}^{\beta_{T}\omega_{j}}+2\Re(\alpha^{*}\beta)\right)$,
the fluctuating trajectory of the Hamiltonian has the same probability.
With this in mind, we can interpret the single trajectory as the action
of the environment (being in state $\bm{\alpha}$) on the system through
$z(t)$ (and $w(t)$) continuously. Hence, the state of the bath,
defined by $\bm{\alpha}$ (and $\bm{\beta}$) defines the whole time
trajectory. This phenomenon reflects the concept of non-locality of
quantum mechanics. 

It should be denoted that the interpretation described in the previous
paragraph gives a false impression as if the bath does not respond
to the system dynamics. This indeed follows from $\kappa^{1}$ approach
as the Hamiltonian is linear to Gaussian fluctuations $z(t)$ and
$w(t)$. The picture is slightly different as we consider terms $\kappa^{2}$
or higher (Eq. \eqref{eq:gen stoch schr eq}). With $\kappa^{2}$
terms the time-dependent Hamiltonian is also affected by the correlation
function of the fluctuations, hence the relaxation of the bath with
respect to system state is then included up to $\kappa^{2}$. The
$\kappa^{2}$ terms, hence, reflect the back-action, which becomes
non-local and contains the memory of the action. The higher orders
of the action by environment could be included by expanding the non-local
exponential function in Eq. \eqref{eq:gen stoch schr eq}. 

These properties ensure that the dynamics of a composite system+environment
system is treated in the correlated fashion, i. e., the system is
affected by the environment and the environment is affected by the
the system dynamics. These properties allow to describe correlated
dissipative properties, such as for instance polaronic effects, which
appear to be quite important for molecular excitations. The polaron
formation effect has been captured from the equilibrium density matrix
using the HEOM \cite{PhysRevB.84.245430}. The effective resonant
coupling has been found to depend exponentially on the system-bath
coupling strength. The time evolution of this relaxation has been
revealed using the variational approach \cite{chorosajev2013} and
various polaron formation scenarios have been obtained. The SSE could
be used to reveal the dynamical Hamiltonian as well, thus these effects
are captured in the excitation dynamics presented in this paper. Additionally
the SSE includes the coherent effects such as high-frequency molecular
vibrations through resonances of the spectral densities. While the
spectral density approach has limitations in the exciton basis (which
must be fluctuating due to bath dynamics and the cumulant expansion
is not valid), we avoid problems as we treat the problem in the site
basis, which is independent of the fluctuations. Hence the resonances
of the spectral density, reflecting the intra-molecular vibrations
are properly included.

The exact treatment of the open quantum system dynamics at the level
of HEOM requires solving the general form of the SSE (Eq. \eqref{eq:gen stoch schr eq}).
However, we have showed that time-local SSE (Eq. \ref{eq:schr eq with Ht}),
obtained by making second order to the system-bath interaction approximation
guarantees quite accurate results at room temperature. Additionally,
possibility to renormalize the wave-function at an arbitrary time
allows to avoid problems of divergencies. Consequently we can see
from Fig. \ref{fig:heom sse redf compar} that the SSE performs considerably
better than the Redfield equation, especially at low temperatures.
Another useful property of the SSE method is that it does not impose
any restrictions on the spectral density of the heat bath - according
to Eq.\eqref{eq:fluct express} fluctuations $u_{n}\left(t\right)$
can be generated with the function $C^{\prime\prime}\left(\omega\right)$
of arbitrary form and, thus, describe various environments of the
system, including the high-frequency intra-molecular vibrations. 

The theory of decoherence \cite{schlosshauer2007decoherence} dictates
that the interaction between the system an environment (effectively
the measurement) causes decoherence in the system and final collapse
of the system state into the so-called preferred state which is uniquely
defined by the \char`\"{}measurement device\char`\"{}. In this interpretation
the dynamics of the open quantum system can be understood in terms
of the continous measurement which damps the system dynamics. Such
approach has been successfully used in describing the quantum dynamics
\cite{bittner1995quantum,bittner1997decoherent,bittner1997quantum}.
We use the concept of measurement at few points as well. First, our
simulation trajectory corresponds to the measurement of the state
of the bath $|\bm{\alpha}\rangle$. Second, we perform measurement
of the excitation position when we define the transfer time. Compared
to result of Refs. \cite{bittner1995quantum,bittner1997decoherent,bittner1997quantum}
the whole ensemble of our trajectories then correspond to decoherence
of system wavevector as the wavevector diverges for different bath
trajectories (see Fig. \ref{fig:popul exp single traj}). 

The stochastic nature of the wave vector enables us to calculate the
stochastic properties of the system and this feature is a big advantage
of the SSE formalism over the reduced density matrix methods. One
of the important properties available from the SSE is the exciton
transfer time distribution. The stochastic transfer times of the exciton
and their distribution are proper quantities in our approach, while
the reduced density matrix formalism do not define them when the processes
are non-exponential. Our procedure of obtaining the stochastic transfer
time from the SSE is based on an assumption of a quantum measurement
of the system state with the $N$-slit-like measurement device in
analogy with the two-slit experiment: the stochastic populations of
the system represent the probabilities to find the exciton on a particular
site and the random number $r$ models the operation of the exciton
detector at one of the ``slits'' \cite{schlosshauer2007decoherence}.
The adequacy of this procedure of exciton transfer time calculation
has been illustrated by applying this method to a two level system
interacting with a nearly Markovian heat bath which means that the
environmental processes are much faster than those in the system.
For a two-level system the waiting time coincides with our definition
of the transfer time. 

The two-site model system presented in this paper allows to validate
various angles of our approach: in the Markovian weak-coupling case
we find all necessary properties of the dynamics consistent with the
theoretical predictions including the proper thermal equilibrium,
correct scaling of transfer rates, as well as proper exponential distributions
of transfer times. The strong-coupling case has non-exponential evolutions,
consistent with exact HEOM approach. 

\begin{figure}
\noindent \begin{centering}
\includegraphics[width=8cm]{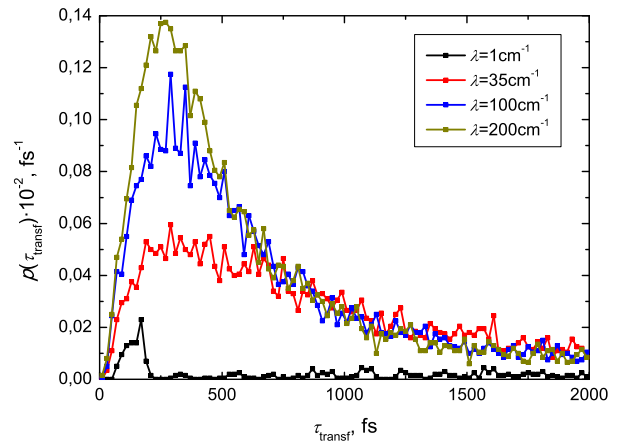}
\par\end{centering}

\caption{\label{fig:transf time lambda} Distributions of energy excitation
transfer time calculated for the 3rd site in the FMO system at different
system - bath interaction strengths. Debye frequency $\omega_{\textrm{D}}=100\textrm{ fs}^{-1}$,
$T=300\textrm{ K}$.}
\end{figure}

However, the main result obtained in this paper is the conclusion
on the effect of intra-molecular high-frequency vibrations on the
energy transfer dynamics in photosynthetic FMO aggregate. Recent 2D
spectroscopy experiments revealed long-lasting quantum coherences
in FMO and a range of other systems \cite{lee2007coherence,Pachon2011,schlau2012elucidation}.
There is a continuous debate on the origin of these beats, while their
assignment recently was shifted to be vibrational. The role of the
coherence is considered to be an important factor for defining the
excitation dynamics in molecular aggregates. We hence addressed the
very core of the problem and simulated the excitation transfer processes
by including or excluding the high-frequency vibrations. As revealed
by the exciton transfer time distribution to the 3rd FMO site shown
in Fig. \ref{fig:transf time lambda}, the overall dynamics certainly
becomes slightly faster, however, the excitation transfer pathways
are not very sensitive to the choice of the spectral density, i. e.,
whether we have or do not have high frequency vibrational modes. It
could be argued that the system-bath coupling strength parameters,
the reorganization energies, of both spectral densities are different
so the results are hardly comarable. However, the reorganization energy,
includes contributions from both the low frequency and the high frequency
components, so obviously the two models of spectral density cannot
have the same reorganization energy. However, the low frequency pars
of the spectral densities are comparable, so the effect on the transfer
times is necesseraly related to the high frequency spectral components.

\noindent 

\begin{acknowledgments}
This work was supported by the Research Council of Lithuania (LMT)
through grant No. MIP 069/2012. We also wish to thank Andrius Gelzinis
for useful discussions.
\end{acknowledgments}
\appendix

\section{\label{sec:coher states}Some properties of coherent state representation}

The coherent state $\left|\alpha\right\rangle $ is defined as the
eigenvector of the annihilation operator $\widehat{a}$:

\begin{equation}
\widehat{a}\left|\alpha\right\rangle =\alpha\left|\alpha\right\rangle .\label{eq:coher state def}
\end{equation}
The annihilation operator is not Hermitian, thus the quantity $\alpha$
in Eq. \eqref{eq:coher state def} is a complex number and can get
any value. The representation of the coherent state $\left|\alpha\right\rangle $
in the energy eigenbasis of the harmonic oscillator $\left|n\right\rangle $
can be obtained by calculating the scalar product of both sides of
Eq. \eqref{eq:coher state def} with the vector $\left\langle n\right|$
which yields

\begin{equation}
\left|\alpha\right\rangle =C\overset{\infty}{\underset{n=0}{\sum}}\frac{\alpha^{n}}{\sqrt{n!}}\left|n\right\rangle .\label{eq:coher state energ eigenst}
\end{equation}
Choosing $C$ to be equal to 1 we can calculate using Eq. \eqref{eq:coher state energ eigenst}
the scalar product of two coherent states $\left\langle \alpha\right|$
and $\left|\beta\right\rangle $

\begin{equation}
\left\langle \alpha|\beta\right\rangle =\textrm{e}^{\alpha^{*}\beta}.\label{eq:scal prod coher}
\end{equation}
From Eq. \eqref{eq:scal prod coher} we can see that the coherent
states $\left|\alpha\right\rangle $ are not normalized because $\left\langle \alpha|\alpha\right\rangle =\textrm{e}^{|\alpha|^{2}}\neq1$.
Despite the fact that the coherent states $\left|\alpha\right\rangle $
being the eigenvectors of a non-Hermitian operator $\widehat{a}$
are not orthogonal, they still can be used to construct the identity
operator

\[
\widehat{\boldsymbol{1}}=\int\frac{\textrm{d}^{2}\alpha}{\pi}\textrm{e}^{-|\alpha|^{2}}\left|\alpha\right\rangle \left\langle \alpha\right|,
\]
where $\textrm{d}^{2}\alpha\equiv\textrm{d}\left[\textrm{Re}\alpha\right]\textrm{d}\left[\textrm{Im}\alpha\right]$
and the factor $\textrm{e}^{-|\alpha|^{2}}$ ensures the proper normalization
\cite{gazeau2009coherent}.

Another useful property of the coherent states can be obtained by
noticing that the vector $\left\langle \alpha\right|\widehat{a}$
can be written as $\frac{\partial}{\partial\alpha^{*}}\left\langle \alpha\right|$:

\begin{eqnarray}
 &  & \left\langle \alpha\right|\widehat{a}=\overset{\infty}{\underset{n=0}{\sum}}\frac{\left(\alpha^{*}\right)^{n}}{\sqrt{n!}}\left\langle n\right|\widehat{a}\nonumber \\
 &  & =\overset{\infty}{\underset{n=0}{\sum}}\frac{\left(n+1\right)\left(\alpha^{*}\right)^{n}}{\sqrt{\left(n+1\right)!}}\left\langle n+1\right|=\frac{\partial}{\partial\alpha^{*}}\left\langle \alpha\right|.\nonumber \\
\label{eq:annih op change}
\end{eqnarray}
Using this property we can deal with the problem of extraction of
the system wave vector from the third term in Eq. \eqref{eq:stoch schro eq first}:

\begin{eqnarray}
 &  & \textrm{e}^{-\alpha^{*}\beta}\left\langle \boldsymbol{\alpha}\right|\widehat{U}\left(t-\tau\right)\widehat{L}_{m}\widehat{U}^{\dagger}\left(t-\tau\right)\widehat{U}\left(t\right)\left|\boldsymbol{\beta}\right\rangle \left|\phi\right\rangle \nonumber \\
 &  & =\hat{T}\textrm{e}^{-i\overset{t}{\underset{\tau}{\int}}\textrm{d}\tau^{\prime}\widehat{\mathcal{H}}_{\alpha}\left(\tau^{\prime}\right)}\widehat{L}_{m}\hat{T}\textrm{e}^{i\overset{t}{\underset{\tau}{\int}}\textrm{d}\tau^{\prime}\widehat{\mathcal{H}}_{\alpha}\left(\tau^{\prime}\right)}\left|\psi\bm{\alpha}_{\phi\bm{\beta}}\left(t\right)\right\rangle .\nonumber \\
\label{eq:non loc term}
\end{eqnarray}
Here we used the expression of the evolution operator $\widehat{U}\left(t\right)=\hat{T}\textrm{e}^{-i\overset{t}{\underset{0}{\int}}\textrm{d}\tau^{\prime}\widehat{H}\left(\tau^{\prime}\right)}$
in the interaction representation, which involves the time-ordering
operator $\hat{T}$ and $\textrm{e}$ denotes the exponential series.
The operator in the exponent is 
\begin{eqnarray}
 &  & \widehat{\mathcal{H}}_{\bm{\alpha}}\left(t\right)=\widehat{H}_{S}\nonumber \\
 &  & +\kappa\underset{n}{\sum}\underset{j}{\sum}\left[\widehat{L}_{n}g_{nj}\textrm{e}^{i\omega_{j}t}\alpha_{j}^{*}+\widehat{L}_{n}^{\dagger}g_{nj}^{*}\textrm{e}^{-i\omega_{j}t}\frac{\partial}{\partial\alpha_{j}^{*}}\right].\nonumber \\
\label{eq:op H_a def}
\end{eqnarray}
Also we introduce a new operator $\widehat{A}_{\bm{\alpha}}\left(t-\tau\right)$:
\begin{eqnarray}
 &  & \widehat{A}_{\bm{\alpha}}\left(t-\tau\right)=\hat{T}\textrm{e}^{-i\overset{t}{\underset{\tau}{\int}}\textrm{d}\tau^{\prime}\widehat{\mathcal{H}}_{\alpha}\left(\tau^{\prime}\right)}.\label{eq:oper A defin}
\end{eqnarray}
This expression is used in Eq. \eqref{eq:gen stoch schr eq}.

\section{\label{sec:Equilibrium-density-operator}Equilibrium density operator
in the coherent state representation}

It can be shown that every operator $\widehat{O}$ which has square
integrable matrix elements $\left\langle -\alpha\left|\widehat{O}\right|\alpha\right\rangle $
can be expressed in the coherent state basis as a diagonal operator
\cite{PhysRevLett.18.752}:

\begin{equation}
\widehat{O}=\int\frac{\textrm{d}^{2}\beta}{\pi}\varphi(\beta)\left|\beta\right\rangle \left\langle \beta\right|,\label{eq:oper coher diag}
\end{equation}
where the function $\varphi(\beta)$ is equal to

\begin{equation}
\varphi(\beta)=\int\frac{\textrm{d}^{2}\alpha}{\pi}\left\langle -\alpha\left|\widehat{O}\right|\alpha\right\rangle \textrm{e}^{\beta\alpha^{*}-\beta^{*}\alpha}.\label{eq:func fi}
\end{equation}
Both expressions \eqref{eq:oper coher diag} and \eqref{eq:func fi}
are also valid in the multidimensional case with $\left|\boldsymbol{\alpha}\right\rangle $
and $\left|\boldsymbol{\beta}\right\rangle $. 

To calculate the bath equilibrium density operator $\widehat{\rho}_{B}$
in the coherent state basis we must first calculate the matrix element
$\left\langle -\boldsymbol{\alpha}\left|\widehat{\rho}_{B}\right|\boldsymbol{\alpha}\right\rangle $:
\begin{eqnarray}
 &  & \left\langle -\boldsymbol{\alpha}\left|\widehat{\rho}_{B}\right|\boldsymbol{\alpha}\right\rangle =Z^{-1}\left\langle -\boldsymbol{\alpha}\left|\textrm{e}^{-\beta_{T}\widehat{H}_{B}}\right|\boldsymbol{\alpha}\right\rangle \nonumber \\
 &  & =\overset{\infty}{\underset{j=1}{\prod}}\frac{1}{\overline{n}_{j}+1}\exp\left(-|\alpha_{j}|^{2}\textrm{e}^{-\beta_{T}\omega_{j}}\right).\label{eq:matr elem ro_t coher}
\end{eqnarray}
where $\overline{n}_{j}=\left(\textrm{e}^{\omega_{j}\beta_{T}}-1\right)^{-1}$
is the Bose - Einstein function. Now the function $\varphi(\boldsymbol{\beta})$:
\begin{eqnarray}
 &  & \varphi(\boldsymbol{\beta})=\int\frac{\textrm{d}^{2}\boldsymbol{\alpha}}{\pi}\left\langle -\boldsymbol{\alpha}\left|\widehat{\rho}_{T}\right|\boldsymbol{\alpha}\right\rangle \textrm{e}^{\beta\alpha^{*}-\beta^{*}\alpha}\nonumber \\
 &  & =\overset{\infty}{\underset{j=1}{\prod}}\frac{1}{\overline{n}_{j}}\exp\left(-|\beta_{j}|^{2}\textrm{e}^{\beta_{T}\omega_{j}}\right).\label{eq:func fi ro_t}
\end{eqnarray}
Finally, using the expression \eqref{eq:func fi ro_t} and according
to Eq. \eqref{eq:oper coher diag} we can write the equilibrium density
operator of the heat bath
\begin{equation}
\widehat{\rho}_{T}=\overset{\infty}{\underset{j=1}{\prod}}\int\frac{\textrm{d}^{2}\beta_{j}}{\pi}\frac{1}{\overline{n}_{j}}\exp\left(-|\beta_{j}|^{2}\textrm{e}^{\beta_{T}\omega_{j}}\right)\left|\beta_{j}\right\rangle \left\langle \beta_{j}\right|.\label{eq:equil dens op coher}
\end{equation}

\end{document}